\documentclass[12pt,preprint]{aastex}
\newcommand{\be}{\begin{equation}}
\newcommand{\ee}{\end{equation}}

\newcommand{\msun}{{$M_{\odot}$~}}
\newcommand{\ledd}{$L_{\rm Edd}$~}
\newcommand{\mic}{$\mu$m }

\newcommand{\gtsima}{$\; \buildrel > \over \sim \;$}
\newcommand{\ltsima}{$\; \buildrel < \over \sim \;$}
\newcommand{\prosima}{$\; \buildrel \propto \over \sim \;$}
\newcommand{\gsim}{\lower.5ex\hbox{\gtsima}}
\newcommand{\lsim}{\lower.5ex\hbox{\ltsima}}
\newcommand{\simgt}{\lower.5ex\hbox{\gtsima}}
\newcommand{\simlt}{\lower.5ex\hbox{\ltsima}}
\newcommand{\simpr}{\lower.5ex\hbox{\prosima}}

\newcommand{\ie}{{i.e.~}}
\newcommand{\etal}{{et al.~}}
\newcommand{\cxo}{\textit{Chandra~}}
\newcommand{\spi}{\textit{Spitzer~}}
\newcommand{\4}{{V404~Cyg}}

\shorttitle{SED of quiescent black hole X-ray binaries}
\shortauthors{Gallo et al.}

\begin{document}

\title{The Spectral Energy Distribution of Quiescent Black Hole X-ray Binaries: 
New Constraints from \textit{Spitzer}}

\author{Elena Gallo\altaffilmark{1,2},
Simone Migliari\altaffilmark{3}, Sera Markoff\altaffilmark{4}, John A. 
Tomsick\altaffilmark{5}, Charles D. Bailyn\altaffilmark{6}, Stefano
Berta\altaffilmark{3,7}, Rob Fender\altaffilmark{8}, James C. A. Miller-Jones\altaffilmark{4}}
\altaffiltext{1}{Physics Department, Broida Hall, University of
California Santa Barbara, CA 93106} 
\altaffiltext{2}{Chandra Fellow}
\altaffiltext{3}{Center for Astrophysics and Space Sciences, 9500 Gilman Dr.,
University of California San Diego, La Jolla, CA 92093}
\altaffiltext{4}{Astronomical Institute `Anton Pannekoek', University of
Amsterdam, Kruislaan 403, 1098 SJ, Amsterdam, NL} 
\altaffiltext{5}{
Space Sciences Laboratory, 7 Gauss Way, University of California
Berkeley, CA 94720} 
\altaffiltext{6}{Department of Astronomy, Yale University, P.O. Box 208101,
New Haven, CT 06520} 
\altaffiltext{7}{Dipartimento di Astronomia, Universit{\` a} di Padova, Vicolo
dell'~Osservatorio 2, 35122 Padova, IT}
\altaffiltext{8}{School of Physics and Astronomy, University of Southampton,
Southampton SO17 1BJ, UK}

\begin{abstract}
Among the various issues that remain open in the field of accretion
onto black hole X-ray binaries (BHBs) is the way the gas accretes at
very low Eddington ratios, in the so-called quiescent regime. While
there is general agreement that the X-rays are produced by a
population of high-energy electrons near to the BH, the controversy
comes about in modeling the contribution from inflowing vs. outflowing
particles, and their relative energy budget. Recent \spi observations
of three quiescent BHBs have shown evidence for excess emission with
respect to the Rayleigh-Jeans tail of the companion star between 8--24
$\mu$m. We suggest that synchrotron emission from a partially
self-absorbed outflow might be responsible for the observed mid-IR
excess, in place of, or in addition to, thermal emission from
circumbinary material. If so, then the jet synchrotron luminosity, integrated from
radio up to near-IR frequencies, exceeds the measured 2-10 keV
luminosity by a factor of a few in these systems. In turn, the
mechanical power stored in the jet exceeds the bolometric X-ray
luminosity at least by 4 orders of magnitude.  We then compile the
broadband spectral energy distribution (SED) of A0620--00, the lowest
Eddington-ratio stellar mass BH with a known radio counterpart, by
means of simultaneous radio, optical and X-ray observations, and the
archival \spi data. We are able to fit the SED of A0620--00 with a
`maximally jet-dominated' model in which the radio through the soft
X-rays are dominated by synchrotron emission, while the hard X-rays
are dominated by inverse Compton at the jet base. The fitted
parameters land in a range of values that is reminiscent of the
Galactic Center super-massive black hole Sgr A*. Most notably, the
inferred ratio of the jet acceleration rate to local cooling rates is
two orders of magnitude weaker with respect to higher luminosity, hard
state sources.

\end{abstract}

\section{Introduction}
The \textit{Spitzer Space Telescope~}offers the opportunity for the first time
to identify and characterize the properties of highly sub-Eddington Galactic
black hole X-ray binaries (BHBs) in the mid-infrared band, a frequency window
that is still largely unexplored for these systems, and that can prove to be
crucial for our understanding of the overall structure of the accretion flow
in quiescence.  
The infrared (IR) spectra of BHBs with a low mass donor star are
likely shaped by a number of competing emission mechanisms, among
which: reprocessing of accretion-powered X-ray and ultraviolet photons,
either by the donor star surface or by the outer accretion disk,
direct thermal emission from the outer disk, non-thermal synchrotron
emission from a relativistic outflow and thermal emission from
circumbinary dust.
We refer the reader to Russell \etal (2006; R06 hereafter), and
references therein, for a recent comprehensive work on the optical and
near-IR spectral properties of X-ray binaries.  Here we wish to stress
that, as well as for other wavebands, the relative strength of each
mechanism is known to vary greatly in response to changes in the
`X-ray state' of the system (see McClintock \& Remillard 2006; Homan
\& Belloni 2005). Throughout this work, we shall focus on the IR
properties of hard and quiescent low mass BHBs. Such (generally
transient) systems are characterized by strong variability, power-law
dominated X-ray spectra, and integrated X-ray luminosities that are
largely sub-Eddington (roughly between a few $10^{-6}-10^{-2}$ times
the Eddington luminosity, $L_{\rm Edd}$, for the hard state, and below a few
$10^{-6}$\ledd for the quiescent state).

In spite of the large degree of uncertainty on the overall geometry of
the accretion flow in this regime, there is general agreement that the
X-rays are produced by a population of high-energy electrons near to
the BH, and that the accreting gas is highly inefficient at radiating,
either as a result of an intrinsically reduced radiative efficiency
(Narayan \& Yi 1994), or because of a substantial mass loss (Blandford
\& Begelman 1999), or a combination of the two (e.g. Markoff \etal
2001; Yuan \etal 2005).  The hard state is associated with the
production of persistent, partially self-absorbed,
synchrotron-emitting outflows with flat/inverted radio-mm spectra
(Fender 2001). Such jets appear to survive down to quiescent X-ray
luminosities (Gallo \etal 2006), even though sensitivity limitations
on current radio telescopes make it extremely difficult to reach the
signal-to-noise ratios required to assess their presence for low
luminosity systems farther than 2 kpc or so.
There is evidence from large-scale structures that the jets' mechanical power
is comparable to the bolometric X-ray luminosity in some hard state sources
(e.g. Cyg X-1, Gallo \etal 2005a; Russell \etal 2007). However,
even for the highest quality spectral energy distribution (SED), disentangling
the relative contributions of inflow vs. outflow to the radiation spectrum and
global accretion energy budget can be quite challenging, as illustrated by the emblematic
case of 
\object{XTE J1118+480} in McClintock \etal (2003) and Markoff \etal (2001).
Estimates of the total jet power based on its radiation spectrum depend crucially
on the assumed frequency at which the flat, partially self-absorbed spectrum
turns and becomes optically thin, as the jet `radiative efficiency' depends
ultimately on the location of the high-energy cutoff induced by the higher
synchrotron cooling rate of the most energetic particles. Once again, this
quantity has proved hard to measure.

R06 have collected all the available quasi-simultaneous optical and near-IR
data of a large sample of Galactic X-ray binaries over different X-ray
states. The optical/near-IR luminosity of hard/quiescent BHBs correlates with
the X-ray luminosity to the power $\sim$0.6, consistent with the known
radio/X-ray correlation slope down to $10^{-8}$\ledd (Gallo \etal 2006; but
see Gallo 2007 and Xue \& Cui 2007).  Combined with the fact that the near-IR
emission is largely suppressed in the thermal-dominant state (R06, Figure 4),
this leads to the conclusion that, for the BHBs, the break to the optically
thin portion would take place in the mid-IR (2-40 $\mu$m). Additional evidence
for a synchrotron contribution to the IR band in hard state BHBs comes from
variability studies during outbursts (e.g. Hynes \etal 2006; Homan \etal
2005).
Indeed, from a theoretical point of view, the {\it break frequency},
here defined as the frequency at which the partially self-absorbed jet
becomes optically thin, is inversely proportional to the BH mass: as
jet spectral breaks are often observed in the GHz/sub-mm regime in
active nuclei, they are expected to occur in the IR-optical band for
$10^{5-7}$ times lighter objects (see discussion in e.g. Markoff
\etal 2001 and references therein).  
We know however from observations of \object{GX~339--4}, the only BHB
where the optically thin jet spectrum has been perhaps observed
(Corbel \& Fender 2002; Homan \etal 2005), that the exact break frequency can
vary with the overall luminosity, possibly reflecting changes in the
magnetic field energy density, particle density and mass loading at
the jet base (Nowak \etal 2005). Determining the location of
the jet break as a function of the bolometric luminosity is important
to assess the synchrotron contribution to the hard X-ray band, and may 
even highlight substantial differences among different classes of objects.
As an example, the fact that the optically thin jet IR-emission in GX~339--4
connects smoothly with the hard X-ray power law has led to challenge the
`standard' Comptonization scenario for the hard X-ray state of BHBs (Markoff
\etal 2001). On the contrary, recent \spi observations of the ultra-compact
{\it neutron star} X-ray binary 4U~0614+091 (while in a hard state) revealed
that the break frequency must take place in the far-IR in this system,
effectively ruling out a synchrotron origin for the X-ray power law (Migliari
\etal 2006).

In addition to the jet, \spi observations of quiescent BHBs should be
sensitive to possible emission from circumbinary material.
Circumbinary disks may be formed as a result of mass outflow from the
accretion disk, and have been invoked as an efficient process for the
removal of orbital angular momentum in addition to gravitational
radiation loss and/or magnetic braking (see Taam \& Spruit 2001 in
the context of cataclysmic variables).  Alternatively, circumbinary
material could be due to the presence of a post-supernova explosion
fall-back disk, as argued in the case of the anomalous X-ray pulsar 4U
0142+61 (Wang \etal 2006).  
Muno \& Mauerhan (2006; MM06 hereafter) report on \spi observations
of four nearby low mass X-ray binaries: three BHBs plus one neutron
star system.  Excess mid-IR emission -- with respect to the
Rayleigh-Jeans tail of the donor blackbody spectrum -- is detected
from two (possibly all three) BH systems; MM06 attribute this bump to
circumbinary dust that is illuminated by the low mass companion star.
This would imply that the optically
thick-to-thin jet break occurs in the mm regime, at much lower
frequencies than, e.g., inferred by R06.

In this paper, we aim to reassess the relative contribution of the various
emission components to the radio/IR/optical spectra of the BHBs A0620--00, V404 Cyg and
XTE J1118+480 while in the quiescent state. 
We first report on the re-analysis of
\spi observations, focusing on the rms estimate in the 24 \mic datasets 
(Section~\ref{spi}), then proceed by examining the SED of each source
(Section~\ref{sed}). 
The origin of the detected mid-IR excess emission is discussed in
Section~\ref{origin}. We finally focus on the broadband SED of A0620--00, a
highly sub-Eddington ($L_{\rm X}/L_{\rm Edd}\simeq 10^{-8}$) BHB for which
we put together previously published radio/X-ray data, the \spi data and new
optical data, all taken in 2005. We discuss the results of fitting the whole
SED by means of a maximally jet-dominated model in Section~\ref{jet}. A summary is given is
Section~\ref{sum}.

\section{\textit{Spitzer} observations}
\label{spi}

\begin{figure}
\vspace{0.4cm}
\includegraphics[angle=0,scale=.85]{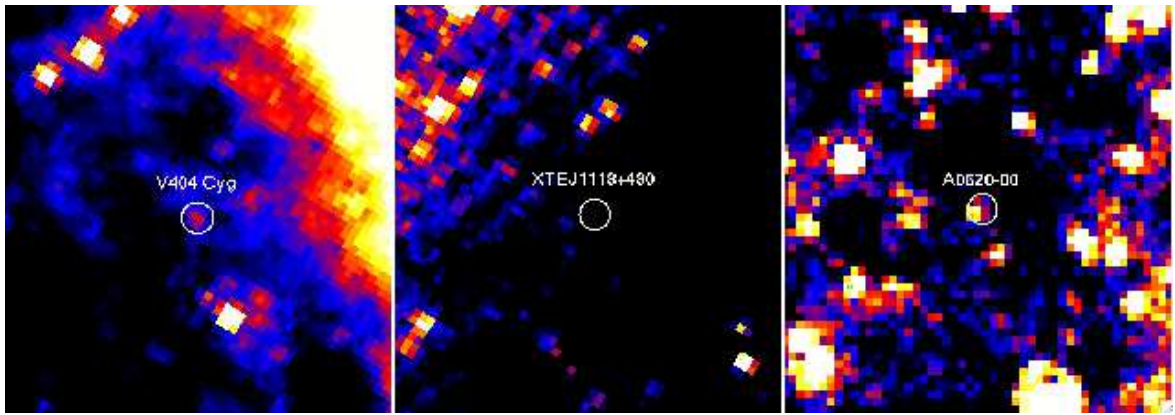}
\vspace{0.4cm}
\caption{
\spi MIPS 24 \mic images of V404 Cyg, XTE J1118+480, and A0620--00. White
circles (with 2 arcsec radius) mark the position of the radio counterparts,
from MERLIN and VLA observations for V404 Cyg (R. Spencer and M. Rupen,
private communications); VLA for A0620--00 (Gallo \etal 2006); MERLIN for XTE
J1118+480 (Fender \etal 2001). The fields of view of V404 Cyg and A0620--00 are
evidently affected by high background contaminations, resulting in high
statistical uncertainties related to sky subtraction. For reference, 1 MIPS pixel corresponds to 1.2 arcsec in size. North is at the top, and East is to the left of these images.
\label{fig:mips}}
\end{figure}

The BHBs A0620--00, V404 Cyg and XTE J1118+480 were observed by {\it Spitzer}
between 2004 October and 2005 May as part of a survey of nearby low-mass
X-ray binaries (PI: Muno, Program 3289).
Photometry of the three targets was acquired using the Multi-band Imaging
Photometer for \spi (MIPS; Rieke \etal 2004) at 24 \mic and the Infrared Array
Camera (IRAC; Fazio \etal 2004) at 8 and 4.5 $\mu$m.
The Basic Calibrated Data (BCD) were re-processed and then mosaicked with the 
{\textsc mopex} software (Makovoz \& Marleau 2005), which uses single, multi-frame,
and dual outlier rejection. 
As discussed by MM06, in the case of A0620--00, the MIPS image was
affected by dark latent features from a previous observation. The
artifacts were corrected by dividing each BCD frame by a normalized
median frame (based on all BCDs excluding the source). These corrected
BCDs were then mosaicked using {\textsc mopex}. Unique IR counter-parts,
consistent with the radio positions, are significantly ($>5\sigma$)
detected at 4.5 and 8 \mic for all the three sources. The MIPS 24 \mic
images of the targets are shown in Figure~\ref{fig:mips}: V404 Cyg and
A0620--00 are detected at the 2-2.5$\sigma$ level, while XTE J1118+480
is undetected.

For each counterpart, we constructed the observation-specific
point-response function (PRF) with {\textsc prf\_estimate}, and extracted
the source flux using both standard aperture photometry on the
background-subtracted image and PRF-fitting (using {\textsc apex}), taking
care to mask foreground stars. 
Sky subtraction
was carried out through the use of multiple 10 arcsec sky apertures placed over an annulus 
around the
source.
Table~\ref{tab:spi} lists the fluxes as measured using both aperture
photometry and PRF-fitting on the mosaic images (the measured fluxes were then
corrected for interstellar extinction following the standard prescription for
the frequency-variable absorption by Cardelli \etal 1989). 
The values obtained with the two methods are consistent with each other within
the errors. While they are also consistent, within the errors, with those
measured by MM06, we derive systematically larger (typically by a
factor 3) rms noise levels for the MIPS 24 \mic fluxes. In fact, statistical
uncertainties related to sky subtraction are usually negligible compared to
calibration and systematic uncertainties. However, statistical uncertainties
can be appreciable -- tens of percent -- for low signal/noise sources
(e.g. Dale \etal 2005). At 24 $\mu$m, this is clearly the case for A0620--00
and V404 Cyg, which are both affected by high cirrus background, as apparent
from Figure~\ref{fig:mips}.
 
%
\section{Radio/Infrared/Optical spectra}
\label{sed}
\begin{figure}
\vspace{0.4cm}
\includegraphics[angle=0,scale=.78]{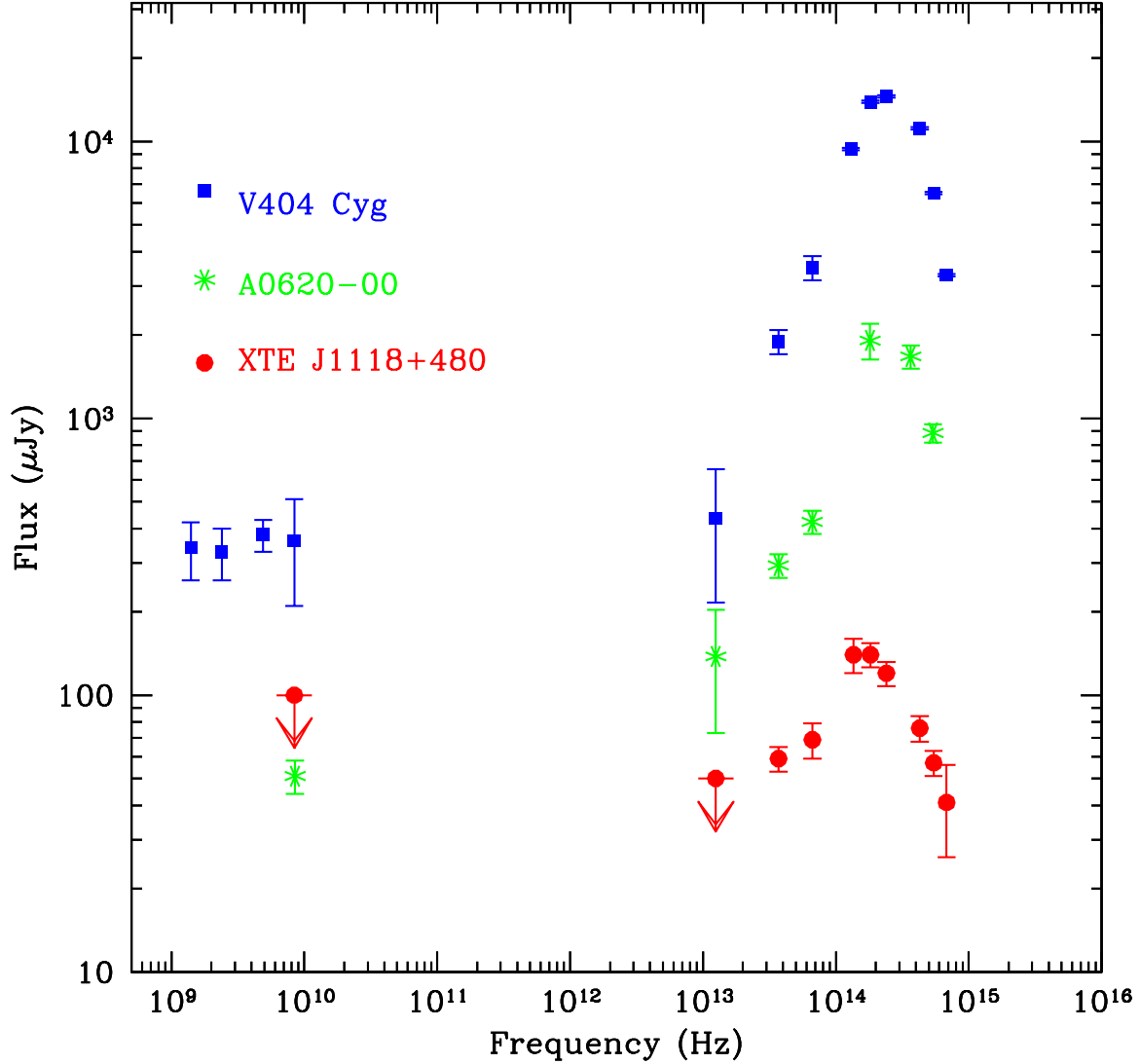}
\vspace{0.4cm}
\caption{Composite radio/IR/optical spectra of quiescent black hole binaries. \textit{V404 Cyg}:
radio data from Gallo \etal (2005b), taken in 2002; IR data from
this work, taken in 2004-2005; optical photometry from Casares \etal 1993,
taken between 1990-1992.
\textit{A0620--00}: radio data 
from Gallo \etal (2006), acquired in August 2005; IR and optical data from
this work. The data span a period of 5 months, with nearly
simultaneous radio/optical coverage. \textit{XTE
J1118+480}: radio upper   
limit from Mirabel \etal (2001); IR data from this work; optical
photometry from Gelino \etal (2006).
\label{fig:seds}}
\end{figure}

We first compile the SEDs of the three systems by putting together the 
\spi data discussed above, plus optical and radio data available in the
literature. For A0620--00, we make use of new optical data, presented in
Section~\ref{a0620sed}. Clearly, the non-strict simultaneity of the observations,
combined with the known variability of quiescent BHBs at all wavelengths
(e.g. Hynes \etal 2003, 2004),
should be kept in mind before drawing any definitive conclusion on the
modeling. Figure~\ref{fig:seds} shows the broadband SEDs of V404 Cyg, XTE
J1118+480 and A0620--00, while in the quiescent state, from radio to optical
wavelengths. 

\begin{figure}
\includegraphics[angle=-90,scale=.22]{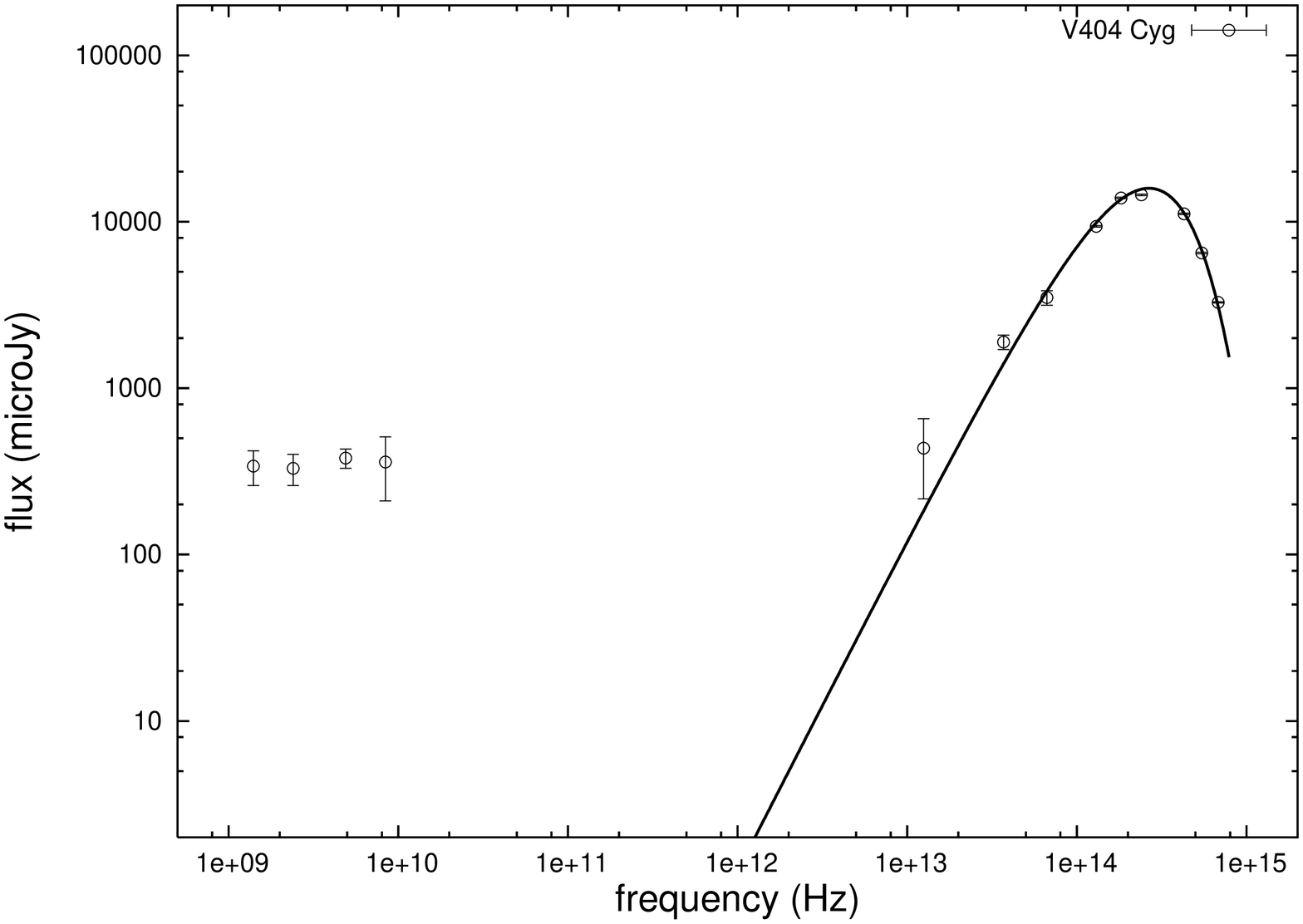}\includegraphics[angle=-90,scale=.22]{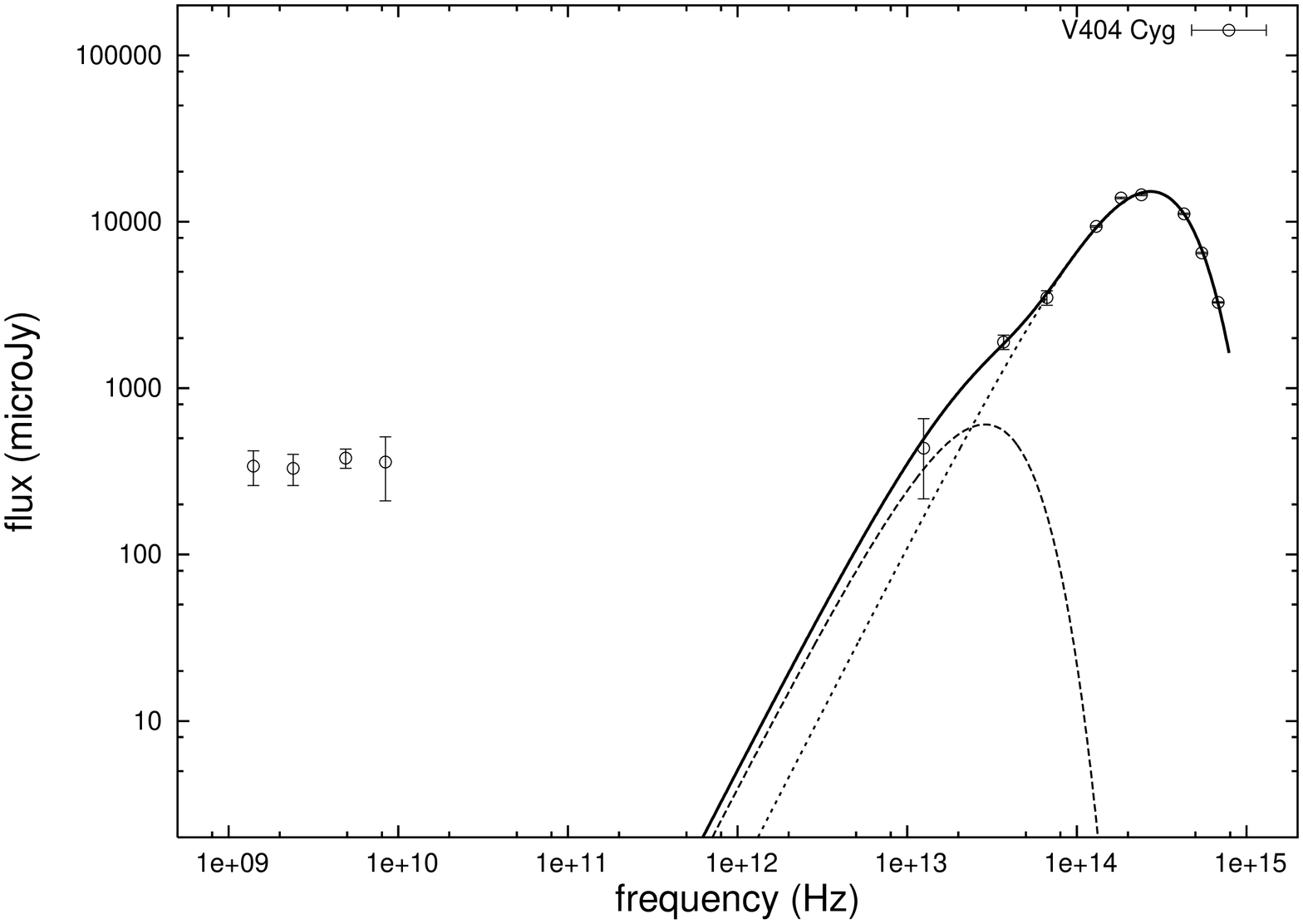}\includegraphics[angle=-90,scale=.22]{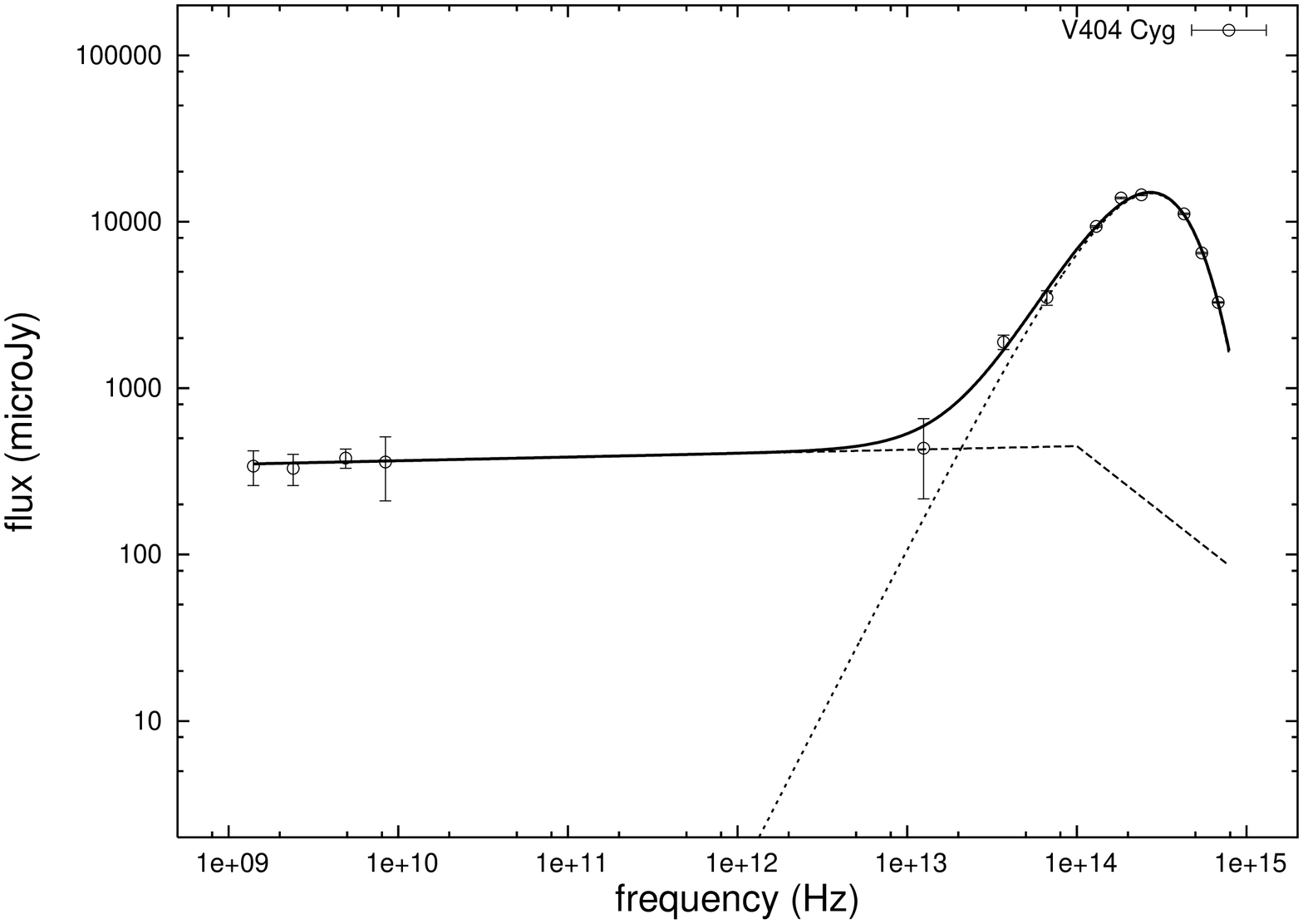}\\
\includegraphics[angle=-90,scale=.22]{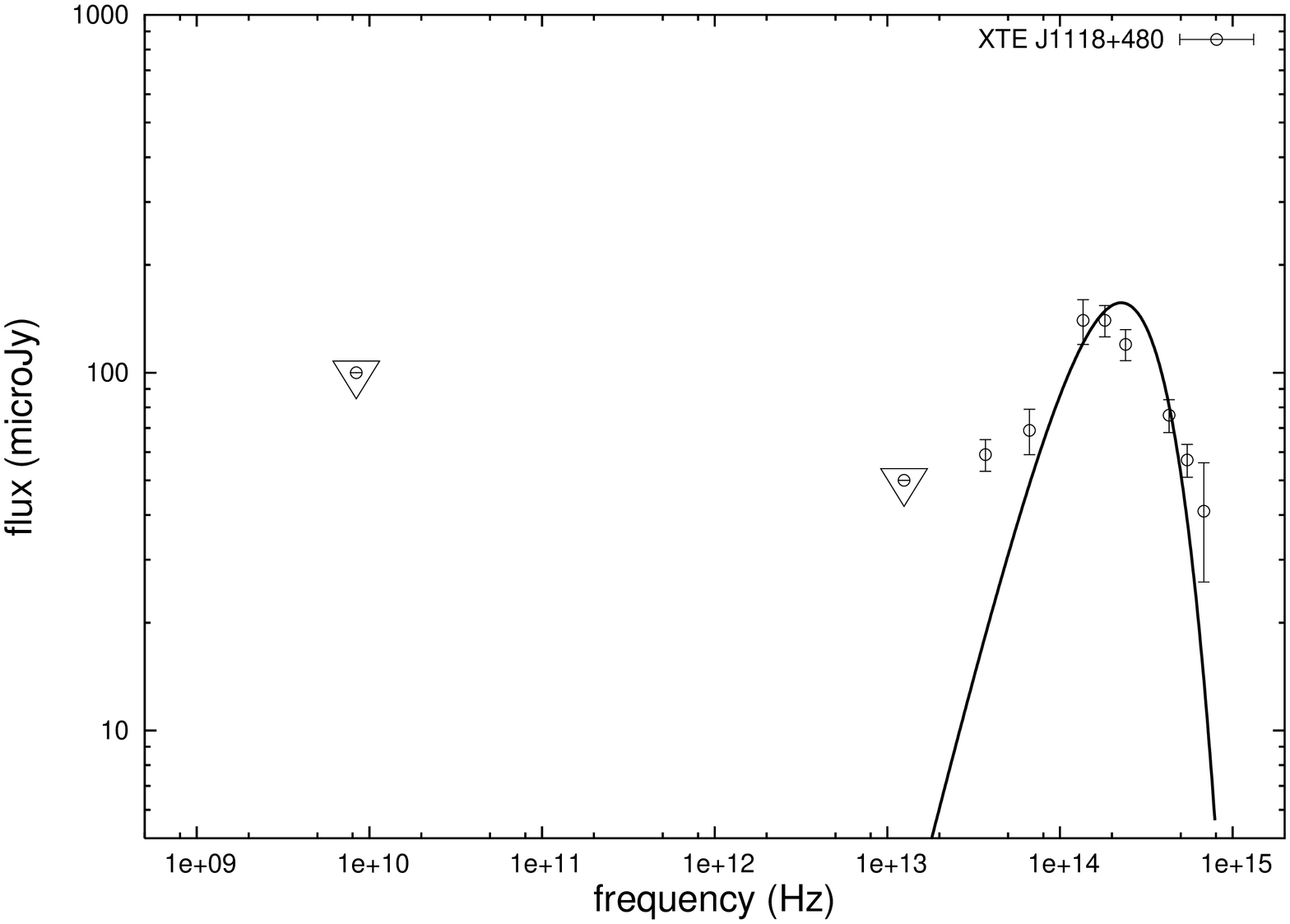}\includegraphics[angle=-90,scale=.22]{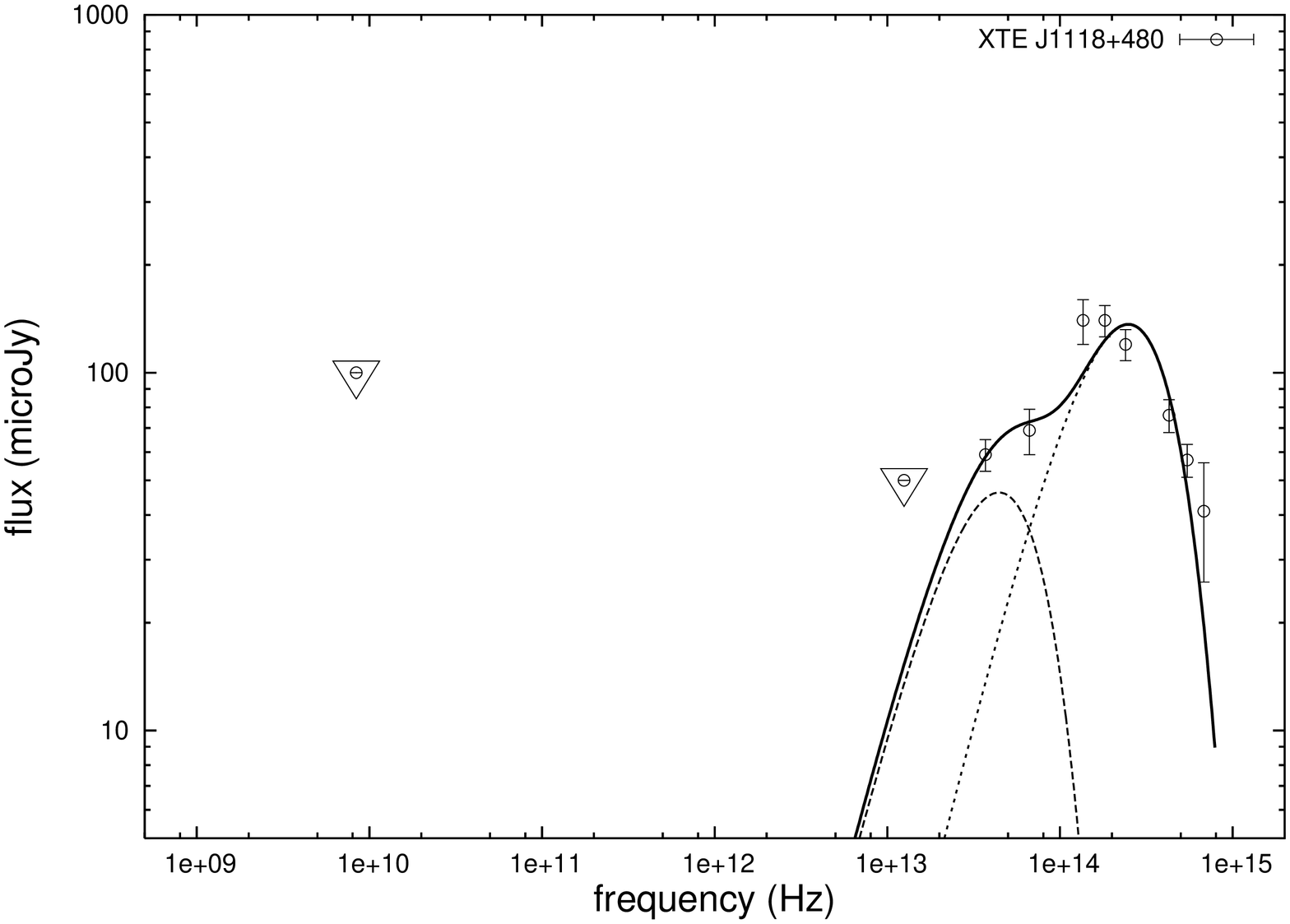}\includegraphics[angle=-90,scale=.22]{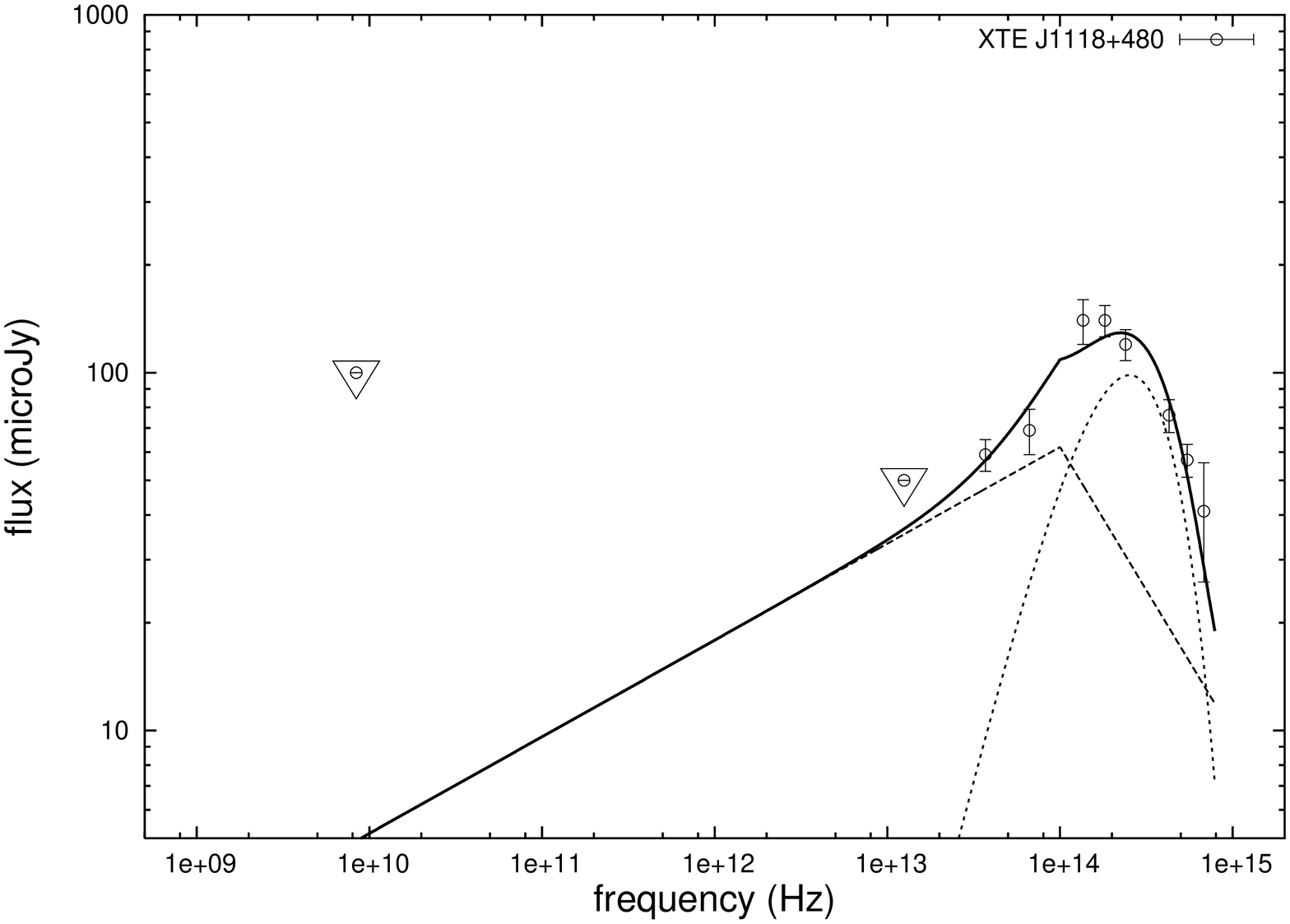}\\
\includegraphics[angle=-90,scale=.22]{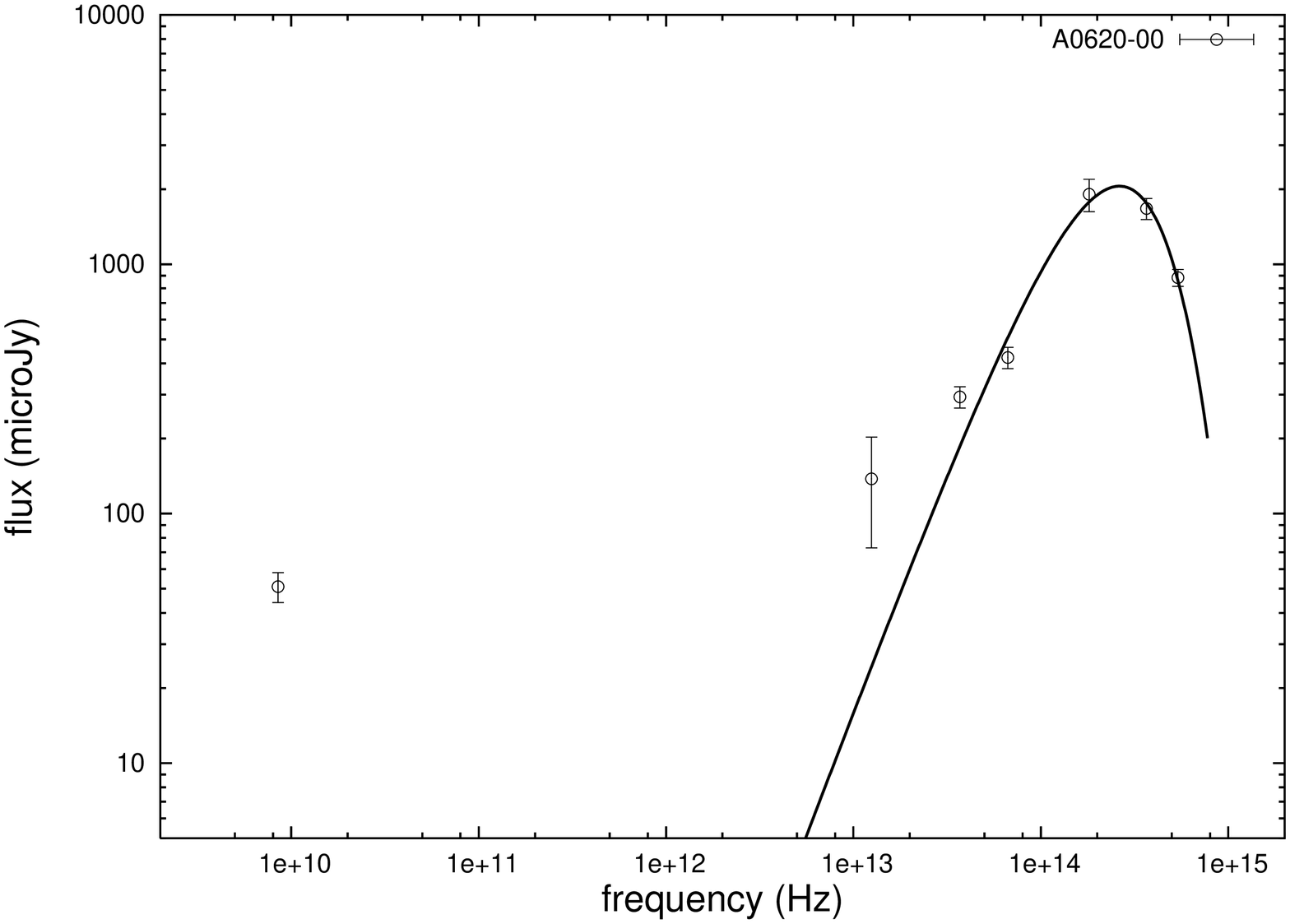}\includegraphics[angle=-90,scale=.22]{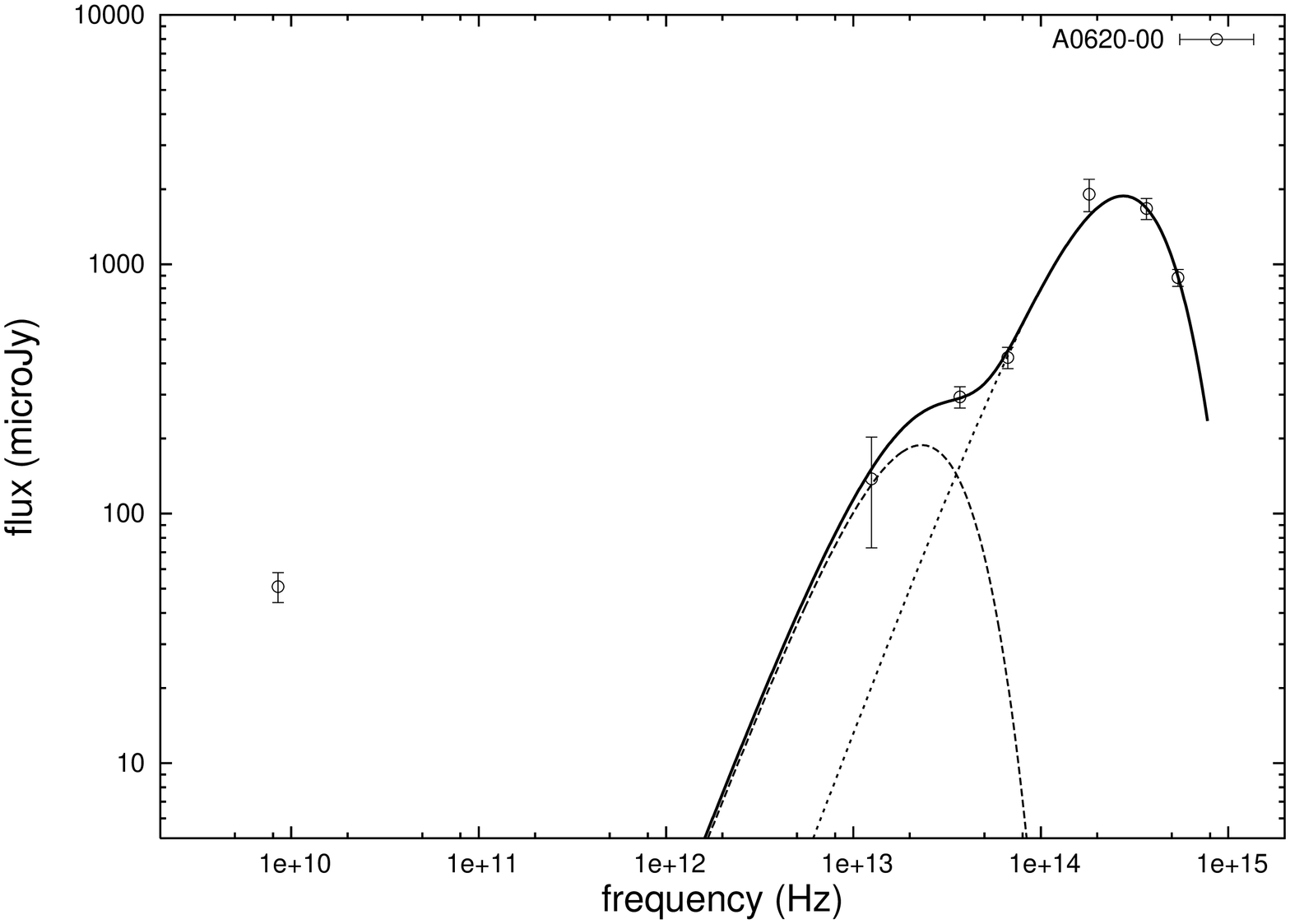}\includegraphics[angle=-90,scale=.22]{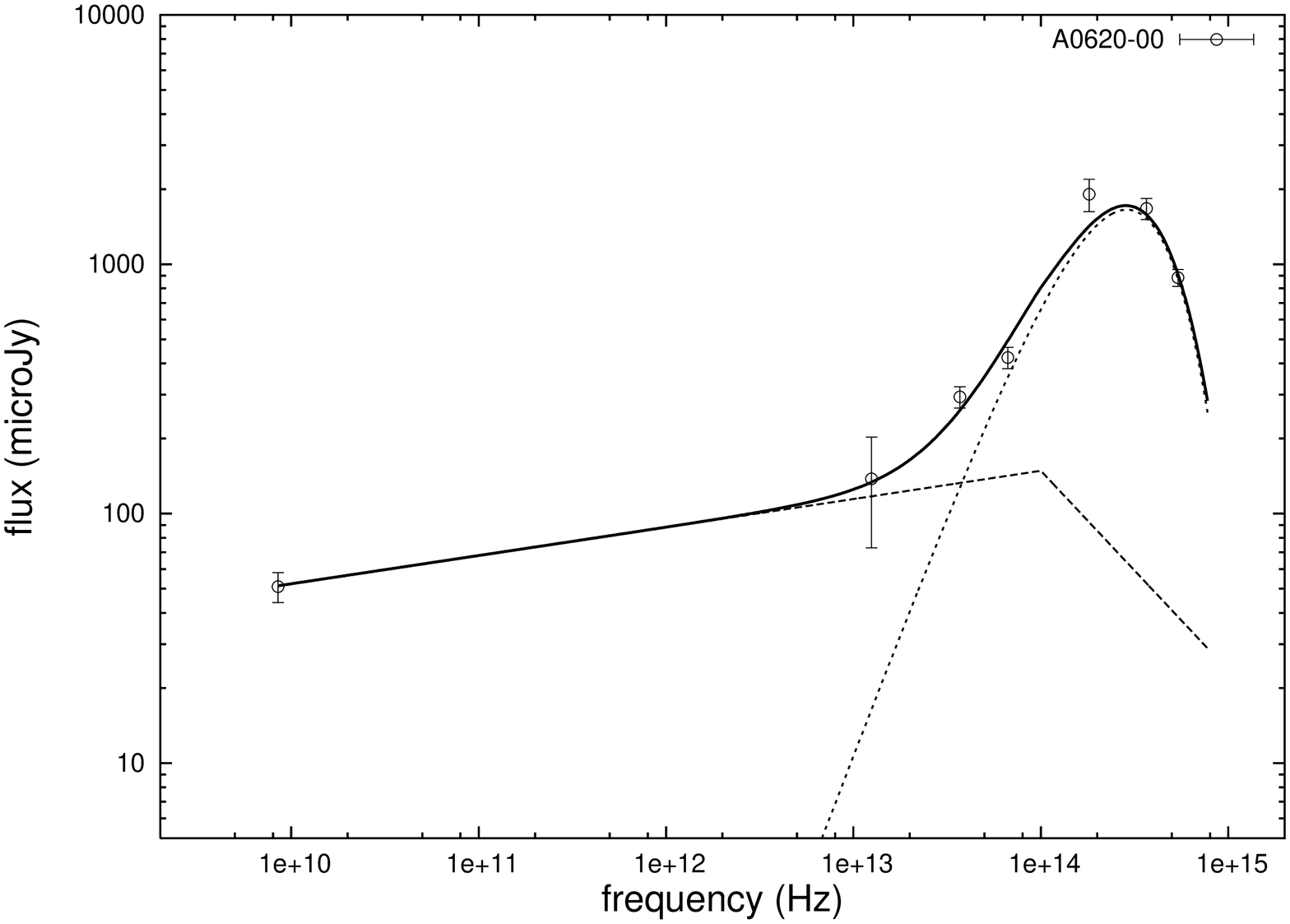}
\vspace{1cm}
\caption{From top to bottom: V404 Cyg, XTE J1118+480, A0620--00. Curves on
the left panels show the fits to the IR/optical data with a single blackbody curve (see
Table~\ref{tab:bb} for the fitted parameters); curves in middle are for
a double blackbody fit (Table~\ref{tab:bb-bb}); the right panels show the
fit to the radio-IR-optical SEDs with a single blackbody plus a broken power law (Table~\ref{tab:bb-pl}).
\label{fig:eg}}
\end{figure}

We first focus on the IR-optical spectra: unlike MM06, we do not compare the
data against stellar atmosphere models: the smoothness of our SEDs does not
demand a sophisticated model which can account for fine spectral features.
Most importantly, we aim to quantify the relative goodness of the various
models via proper $\chi^2$ fitting, which would be meaningless if we were to
apply stellar atmosphere codes to our sparse data-points.  Hence, for each
object we first model the IR-optical spectrum with a single temperature
blackbody, using the best available estimates for the source distance,
inclination and effective temperature. The blackbody approximation is meant to
mimic the contribution from the donor star. As shown by MM06, the
contribution from the irradiated outer accretion disk is negligible for the
parameter space relevant to these quiescent systems, at least in the IR band.
The best-fitting blackbody curves are shown in the left panels of
Figure~\ref{fig:eg}, with the fitted parameters and reduced $\chi^2$ given in
Table~\ref{tab:bb}. Evidently, the single blackbody model provides a poor fit
to the data: excess mid-IR emission, with respect to the Rayleigh-Jeans tail
of the donor/disk, is detected in all three cases.

Fitting the data with two blackbodies (Figure~\ref{fig:eg}, middle panels)
slightly improves the reduced $\chi^2$ in all cases
(Table~\ref{tab:bb-bb}). The temperatures and normalizations of these
secondary blackbodies imply indeed larger physical sizes than the orbital
separation, possibly supporting the circumbinary material interpretation
(MM06).  However, radio emission has been detected in two of these sources
(V404 Cyg: Hjellming \etal 2000, and A0620--00, Gallo \etal 2006), and
interpreted as partially self-absorbed synchrotron emission from a
relativistic outflow. The flat/slightly inverted outflow spectrum must become
optically thin at higher frequencies, possibly in the mid-IR (R06). We thus
explore the possibility that the mid-IR excess might be, at least partly, due
to non-thermal emission from a jet. This possibility has been ruled out by
MM06 on the basis of far too low fluxes/upper limits at 24 $\mu$m. However,
our revised estimates for the 24 \mic rms noise levels leave this possibility
open.

We choose to fit the {\it radio}/IR/optical SEDs with a 
single blackbody plus a broken power law of the form:
\begin{equation}
\label{eq:bp}
F_{\nu}  = F_{\nu_0} \times  \left\{
        \begin{array}{ll}
        (\nu / \nu_0)^{\alpha_1},   &  \nu < \nu_b  \\
        (\nu_b / \nu_0)^{(\alpha_1 - \alpha_2)} (\nu / \nu_0)^{\alpha_2},    & \nu > \nu_b   \\
        \end{array}\right.\;
\end{equation} 
This is meant to account for a partially self-absorbed synchrotron
spectrum with index $\alpha_1=0.0-0.5$ up to the break frequency
$\nu_b$, above which it becomes optically thin with index $\alpha_2$.
After running a grid of models with all the six fitting parameters
(blackbody temperature and normalization, plus the four broken power
law parameters) free to vary, we choose to fix the index of the
optically thin portion to $\alpha_2 = -0.8$ (corresponding to a `canonical' 
electron distribution $N(E)\propto E^{-p}$ with power law index
$p=+2.6$, $E$ being the electron energy; e.g. Fender 2006) and the
position of the break to $\nu_b=10^{14}$ Hz, in order to maximize the
jet contribution to the mid-IR band.  The results of the blackbody plus
broken power law fits are shown in the right panels of
Figure~\ref{fig:eg}, with the fitted parameters in
Table~\ref{tab:bb-pl}.  We discuss below the SED compilation and the
results of the modeling on a case by case basis.
\subsection{V404~Cyg (GS~2023+338)}
Casares \etal (1993) report on $B$-$V$-$R$-$J$-$H$-$K$~band photometry
of \4 taken in 1991 July-August, 2 years after the end of the 1989
outburst that preceded the current quiescent regime (even though this
system, because of its relatively high quiescent X-ray luminosity
[$L_{\rm X}/L_{\rm Edd}\simeq 10^{-6.5}$], is often considered at the
boundary between `quiescence' and the hard X-ray state).  Several
later works have established \4 to be variable by a factor of a few at
IR-to-X-ray wavelengths (see e.g. Hynes \etal 2004, Bradley \etal 2007
for the X-ray/optical variability; Zurita \etal 2004 for a study of
the long term optical/IR variability, and references therein).  The
origin of such variability is yet to be well understood, even though
there is general agreement that it should take place somewhere in the
accretion flow rather than in the hot gas stream/donor star (Shahbaz
\etal 2003; Zurita \etal 2003; Hynes \etal 2003; 2004).

Over the past few years, V404 Cyg has been known as a relatively stable radio
source, with an average flux density of $\sim 350~\mu$Jy, and a flat/slightly
inverted spectrum at GHz frequencies (Hjellming \etal 2000; Gallo \etal 2005b), interpreted in
terms of partially self-absorbed synchrotron radiation from outflowing plasma.
The variable nature of this system, combined with the fact that the available
data spread an interval of several years (the optical and \spi data
were acquired more than 10 years apart), make it especially difficult
to draw definite conclusions about the mid-IR emission detected with
MIPS (on the other hand, R06 showed that the optical-IR luminosity of
hard/quiescent state sources scales with the X-ray luminosity to the
power 0.6, implying that the X-ray variability should be reduced to
some extent in the IR).  

The top panels of Figure~\ref{fig:eg} show the IR-optical spectrum of
V404 Cyg as fitted with a single and double blackbody model (left and
middle panel, respectively): clearly the latter model provides a
better fit to the IR-optical data, with $\chi^2$/d.o.f.=10.3/7 and
$\chi^2$/d.o.f.=1.4/5, respectively.
However, these components do not account for the radio emission.
Because of the flat radio spectrum, it can not be ruled out that the
excess emission at 24 \mic might be due to the high frequency portion
of the well-established synchrotron-emitting outflow.  The top right
panel of Figure~\ref{fig:eg} shows a fit to the radio-IR-optical data
with a single blackbody with $T\simeq 4600$ K plus a broken power law,
where the fitted index of the partially self-absorbed regime is
$\alpha_1=0.02$.  This two-component model provides as a good fit as
the double blackbody model ($\chi^2$/d.o.f.=4.0/9), and it also accounts
for the radio emission.

This suggests that, in this system, synchrotron emission from a
partially self-absorbed outflow is likely to be responsible for the
observed mid-IR excess as much as thermal emission from circumbinary
material.  As an aside, if such excess were entirely due to
circumbinary disk emission, this would imply that the jet break to the
optically thin portion has to occur somewhere in the mm regime, \ie at
lower frequencies than predicted by R06. While the system SED could be
comfortably reproduced by the sum of two blackbody components plus a
broken power law, accommodating the circumbinary material and jet
the contribution, this would require as many free parameters as data points.
\subsection{XTE~J1118+480}

Gelino \etal (2006) present $B$-$V$-$R$-$J$-$H$-$K_S$ band photometry
of XTE~J1118+480 in quiescence.  Due to its high Galactic latitude,
XTE~J1118+480 is a virtually unabsorbed source ($A_V=0.06$) and yet it
can be taken as an example of how tricky it can be to infer the
properties and the geometry of the accretion flow based on modeling
the SED. For instance, the high quality simultaneous multi-wavelength
data acquired while in the hard state (at $L_{\rm X}/L_{\rm Edd}\simeq 10^{-3}$)
have been successfully modeled in terms of an advection-dominated
accretion flow (McClintock \etal 2003; Yuan \etal 2005), as well as
using a jet synchrotron model (Markoff \etal 2001).  As shown in
Figure~\ref{fig:eg}, middle panels, there is evidence for substantial
excess emission at 8 \mic with respect to the donor star tail.  We
notice that the single blackbody model provides a very poor
representation of the donor star spectrum: in this case, the actual
stellar atmosphere model is certainly more appropriate (see MM06,
Figure 2).  However, as noticed above, we are interested in
constraining the nature of the excess mid-IR emission via proper
$\chi^2$ fitting: in this framework, irrespective of how well the
donor star thermal emission is modeled, our goal is to determine
whether fitting the mid-IR excess with a broken power law model
provides a better or worse description of the data in a statistical
sense.  The radio counterpart to XTE J1118+480 is undetected in
quiescence, with an upper limit of 0.1 mJy at 8.5 GHz (Mirabel \etal
2001).  Because of this shallow upper limit, the measured excess at 8
\mic might still be interpreted as due to a partially self-absorbed
outflow that extends its power-law spectrum from the radio up to the
IRAC regime. This is illustrated in middle right panel of
Figure~\ref{fig:eg}, where a partially self-absorbed synchrotron
emitting outflow with $\alpha_1=0.27$, plus a $\sim 4300$ K blackbody
component, account for the system SED.  
The reduced $\chi^2$ is
improved with respect to the double blackbody model ($\chi^2$/d.o.f. =
12.0/3 vs. $6.2/3$, respectively for the double blackbody and
blackbody plus broken power law model). Within the blackbody plus
power-law model, the fitted values for the blackbody temperature and
normalization are consistent, within the errors, with the inferred
values for the donor star (namely $\sim 4250$ K and $\simeq$0.4 $R_{\odot}$;
Gelino \etal 2006).  
The fitted radio spectral index is consistent with hard state sources
(Fender 2001), and predicts a GHz flux density lower than 5 $\mu$Jy,
practically undetectable with current radio facilities over reasonable
integration times (the rms noise level for a 24 hr integration with
the VLA is about 5 $\mu$Jy at 8.5 GHz; however, planned upgrades, such as the
eMERLIN and EVLA will be able to probe such flux density levels in
hrs-long exposures).

\subsection{A0620--00 (V616~Mon)}
\subsubsection{SMARTS observations}\label{a0620sed}
We construct the SED of A0620--00 by means of radio, IR, optical and X-ray
observations, all taken in 2005; the optical/near-IR data were acquired by the
Small and Moderate Aperture Research Telescope System
(SMARTS\footnote{http://www.astro.yale.edu/smarts}) consortium, using the
Cerro Tololo Inter-American Observatory (CTIO) 1.3 m together with
ANDICAM\footnote{http://www.astronomy.ohio-state.edu/ANDICAM}, a dual-channel
imager capable of obtaining optical and IR data simultaneously.
A0620--00 was observed through $I$-$V$-$H$ filters on 2005 August 18, one day
before the beginning of the (strictly simultaneous) Chandra/VLA observations
(taken on 2005 August 19-20; 
Gallo \etal 2006), while the \spi data discussed above were acquired on 2005
March 06 (MIPS) and March 25 (IRAC).

SMARTS data were calibrated using data from previous nights and were processed
and reduced using standard IRAF aperture photometry routines.  The measured
magnitudes were converted into fluxes using the SMARTS photometric
zero-points; we used a color excess of $E(B-V)$=$0.39\pm0.02$ (Wu
\etal 1976), and corrected for extinction following again the standard prescription
for the frequency-variable absorption by Cardelli \etal (1989).  
The results are summarized in Table~\ref{tab:smarts}.
Interestingly, all of the measured magnitudes are brighter than the maximum
magnitude from the previously published quiescent light-curves (see Table 1 in
Gelino \etal 2001, reporting on optical and IR observations of A0620--00
between 1976 and 2001), and from 0.5-0.7 mag brighter than the mean magnitudes.
However, given the observed trend over the past few years of increasing
brightness in this source, it seems very unlikely that these results
require a sudden flare. This however has to be kept in mind when inspecting
the whole SED of A0620--00, in particular when comparing the 2005 March \spi
observations with the optical, near-IR values given by Gelino \etal (2001).

\subsubsection{Broadband SED}
Significant excess emission with respect to the Rayleigh-Jeans 
portion of the donor's blackbody spectrum is detected at 8 and 24 $\mu$m.
As shown in the middle-bottom panel of Figure~\ref{fig:eg}, the sum of two
blackbodies ($\sim 4700 + 390$ K) provides
a good fit to the IR-optical data ($\chi^2$/d.o.f.=2.0/2).  
The detection of a radio counterpart to A0620--00  strongly
suggests that this quiescent system is powering a synchrotron-emitting
outflow (Gallo \etal 2006). Arbitrarily assuming a flat spectrum for the
partially self-absorbed portion of the jet, this would have to become
optically thin at frequencies lower than $10^{13}$ Hz for it not to contribute
to the mid-IR excess. Alternatively, the whole radio-IR-optical spectrum can be well
 fit by
the sum of $\sim 4900$ K blackbody plus a broken power law with slightly inverted
spectrum in the radio-IR regime with $\alpha_1=0.1$ (yielding $\chi^2$/d.o.f.=7.8/3).

\section{Origin of the mid-IR excess: implications for the jet power}
\label{origin}
The \spi observations of three quiescent BHBs discussed above show evidence
for excess emission in the mid-IR band; while it may possible to reproduce the
emission between $2-4\times 10^{14}$ Hz with a blackbody whose temperature is
consistent with the shown temperatures of the secondary stars, it would be
difficult to explain the excess at $10^{13}$ Hz with any model for which the
temperature is high enough so that $10^{13}$ Hz is in the Rayleigh-Jeans
portion of the blackbody spectrum.  Thus, two main possibilities arise to
account for the measured excess: thermal emission from cool (hundreds of K)
circumbinary material, or synchrotron emission from outflowing plasma.  The
latter hypothesis was dismissed by MM06 on the basis of far too low 24 \mic
fluxes/upper limits. Our estimates for the statistical uncertainties on the 24
$\mu$m observations, however, reinstate this
possibility. 
 
Under the assumption that non-thermal synchrotron emission is at the
origin of the measured IR-excess, we can estimate the amount of power
stored in the outflows. Integrating the partially self-absorbed jet
spectra up to $10^{14}$ Hz, and assuming a (conservatively low) jet
radiative efficiency of 5$\%$, and no Doppler boosting (see Fender
2001), we obtain jet powers in the range $\sim 4 \times 10^{32}$
erg s$^{-1}$, for A0620--00 and XTE J1118+480, the lower Eddington
ratio sources, up to $\sim 2 \times 10^{34}$ erg s$^{-1}$, for V404
Cyg (see Table~\ref{tab:jets}). Under these assumptions, the total jet
power exceed the measured X-ray luminosities (between 2-10 keV) in
quiescence by a factor 50 at least. Assuming that the steep X-ray
power laws observed in quiescent BHBs (with average photon index
$\Gamma\simeq 2$; e.g. Corbel et al. 2006) extend up to $\sim 100$
keV, where a spectral cutoff is observable in higher Eddington-ratio
systems, the {\it bolometric} (0.1-100 keV) X-ray luminosities
are likely to exceed the measured 2-10 keV luminosities
by a factor of a few. Therefore, this regime of $L_{\rm j, tot}\simgt
L_{\rm X}$ fits the definition of `jet-dominated' state put forward by
Fender \etal (2003).
The above estimates of $L_{\rm j,tot}$ are based on a conservative
radiative efficiency for the synchrotron process of 5$\%$; as such,
they represent strict lower limits.  Alternatively, we can estimate
the total jet power following the formalism by Heinz \& Grimm (2005),
where the monochromatic radio core emission ($L_{\rm r}$, in units of
$10^{30}$ erg s$^{-1}$) of three well studied radio galaxies was
directly compared to the radio lobe emission, and combined with a
self-similar jet model (Heinz \& Sunyaev 2001) in order to calibrate
the ratio of mechanical vs. radiative power of partially self-absorbed
jets.  They proposed that the jet kinetic power of both super-massive
and stellar size BHs can be estimated from the core radio luminosity
as: $ L_{\rm j,tot} = 6.2 \times 10^{37}L_{\rm r}^{1/(1.4-\alpha_{\rm
r}/3)}{\cal W}_{37.8}$ {erg} {s}$^{-1}$, where $\alpha_{\rm r}$ is the
radio spectral index over the partially self-absorbed regime, and the
parameter ${\cal W}_{37.8}$ carries the (quite large) uncertainty on
the radio galaxy calibration.  The normalization value by Heinz \&
Grimm is roughly in agreement to that estimated by K\"ording \etal
(2006): here, for flat spectrum radio sources, the jet power (at the
hard to soft state transition) is expressed as: $L_{\rm
j,tot}\simlt 3.6\times 10^{37} (f/0.75)(\eta/0.1)L_{\rm r}^{(12/17)}$
erg s$^{-1}$, $f$ being the fraction of outer mass accretion rate that
is not expelled via winds/outflows, and $\eta$ the standard accretion
efficiency.  Either way, the inferred total jet power would exceed
the {\it bolometric} X-ray luminosity by at least 4 orders of
magnitude for the three quiescent BHBs under consideration.
It is worth mentioning that, independently of normalization and
efficiency factors, in all three cases the jet {\it synchrotron} 
luminosity, integrated up to $10^{14}$ Hz (that is neglecting the optically thin
portion), already exceed the measured 2-10 keV luminosities by a factor
of a few (Table~\ref{tab:jets}, right column).
 
In contrast, if thermal emission from circumbinary disk material is entirely
responsible for the measured mid-IR excess, this would imply that the jet
spectrum breaks at much lower frequencies, perhaps in far-IR/mm regime,
lowering the above estimates by a factor of ten at least.  A final test to
assess the origin of the measured excess could be variability study in the
mid-IR, possibly coordinated with the radio.

\section{A maximally jet-dominated model for the quiescent state}
\label{jet}

Ultimately, as discussed by McClintock \etal (2003), while there is
general agreement that the X-ray emission in quiescent BHBs comes from
high-energy electrons near the BH, the disagreement comes about in:
{\it i)} attributing the emission to outflowing vs. inflowing
electrons; {\it ii)} modeling the electron distribution as thermal
vs. non-thermal (or hybrid). The SEDs of quiescent BHBs, as well as
low-luminosity AGN are often examined in the context of the
advection-dominated accretion flow model (ADAF; Narayan \& Yi 1994),
whereby the low X-ray luminosities would be due to a highly reduced
radiative efficiency, and most of the liberated accretion power
disappears into the horizon. Alternatively, building on the work by
Falcke \& Biermann (1995) on AGN jets, a jet model has been proposed
for hard state BHBs. The model is based upon four assumptions: 1) the
total power in the jets scales with the total accretion power at the
innermost part of the accretion disk, $\dot{m}c^2$, 2) the jets are
freely expanding and only weakly accelerated via their own internal
pressure gradients only, 3) the jets contain cold protons which carry
most of the kinetic energy while leptons dominate the radiation and 4)
some fraction of the initially quasi-thermal particles are accelerated
into power-law tails.  Markoff \etal (2001) argued that jet
synchrotron emission could account for the broad continuum features of
the simultaneous radio through X-ray observations of XTE J1118+480
while in the hard state. This same model could also explain the broad
spectral features of 13 quasi-simultaneous radio/X-ray observations of
GX 339--4, and was able to reproduce the observed non-linear
radio/X-ray correlation in this system (Corbel et al. 2003) by varying
the amount of power that is channeled in the jet (Markoff et
al. 2003).  Based on the required reflection signatures a new model
was developed (Markoff \etal 2005; MNW05 hereafter) which could
reproduce the simultaneous radio/X-ray data of hard state systems
(GX~339--4 and Cygnus X-1) via radiation from a compact, mildly
relativistic jet, combined with a truncated thermal disk.  In
particular, the X-ray emission can be interpreted as a combination of
optically thin synchrotron emission predominantly from an acceleration
region $\sim 10-100$ gravitational radii along the jets, plus external
(thermal disk photons) and synchrotron self-Compton emission from the
base of the jets, in a region associated with a magnetic compact
corona.  The radio through the soft X-rays are dominated by
synchrotron emission, while the hard X-rays are mostly Comptonization,
with weak reflection.  This {\it `maximally jet-dominated model'} was
intended to explore the possibility that the `hot
electron corona' and `jet base' may be intimately related, or, in the
extreme case, synonymous (we refer the reader to MNW05 for a fuller
description). This model has been tested extensively on simultaneous
radio and X-ray data, and for a number of hard state BHBs. The mid-IR
portion of the spectrum is clearly crucial in order to put constraints
on the optically thick-to-thin jet breaks, as demonstrated by the \spi
observations of the neutron star X-ray binary 4U 0614+091 (Migliari
\etal 2006) and the BHB GRO 1655--40 (Migliari \etal, submitted to
ApJ).

In the following we attempt to fit the radio through X-rays SED of
A0620--00 in quiescence via the maximally jet-dominated model, where
full details can be found in the Appendix of MNW05. The choice of
A0620--00 (over e.g. V404 Cyg, for which the radio spectrum is well
constrained) is motivated by the fact that, with the exception of the
\spi data, the observations were acquired nearly-simultaneously (the
VLA/\cxo observations were strictly simultaneous, while the SMARTS
observations were taken only one day apart). As a comparison, the
broadband SED of V404 Cyg is built on datasets that were taken over 10
years apart.  In addition, A0620--00 has been in quiescence for over
30 years, and is considered as a stable and moderately variable
system, while V404 Cyg is known to vary in flux by a factor of a few
within hours (e.g. Hynes \etal 2003).

\subsection{Application to A0620--00}
The fitting was performed with the \textsc{Interactive Spectral Interpretation
System} (\textsc{ISIS}; Houck \& De Nicola 2000).  As outlined in MNW05, the
fitting is initiated outside ISIS in order to avoid local minima, using
unfolded data that yield a set of starting parameters for which the reduced
$\chi^2$ is lower than 2.  We have decided to fix several parameters which
previously have been allowed to vary, in some cases because the results of
fitting the model to several hard state sources suggest that there may be
canonical values, and secondly because of the low count rates.  
In spite of the large luminosity difference between A0620--00 ($L_{\rm
X}/L_{\rm Edd}\simeq 10^{-8}$) and other sources whose hard state
spectra were successfully fitted by the jet model, such as XTE
J1118+480 (Markoff \etal 2001), GX339--4 and Cygnus X--1 (Markoff
\etal 2005), simultaneous VLA/\cxo observations of A0620--00 in
quiescence have shown that the non-linear radio/X-ray correlation for
hard state BHBs appears unbroken all the way down to $10^{-8}L_{\rm
Edd}$, arguing for no substantial difference between hard and
quiescent state (Gallo \etal 2006; but see Xue \& Cui 2007 and Gallo
2007). On the other hand, recent high statistics X-ray observations
of hard state BHBs seem to show that a geometrically thin disk is
present and extends close to the innermost stable orbit already at
$10^{-3} L_{\rm Edd}$ (Miller \etal 2006a, 2006b; Rykoff \etal
2007). As such solution would be very difficult to maintain at
$10^{-8}L_{\rm Edd}$, these authors conclude that a major transition
has to take place at intermediate luminosities.  Consequently, in
light of the large degree of uncertainty over the nature and geometry
of the accretion flow in quiescence, this must be considered as an
exploratory study. \\

\begin{figure}
\vspace{0.4cm}
\hspace{0.4cm}
\includegraphics[angle=0,scale=.8]{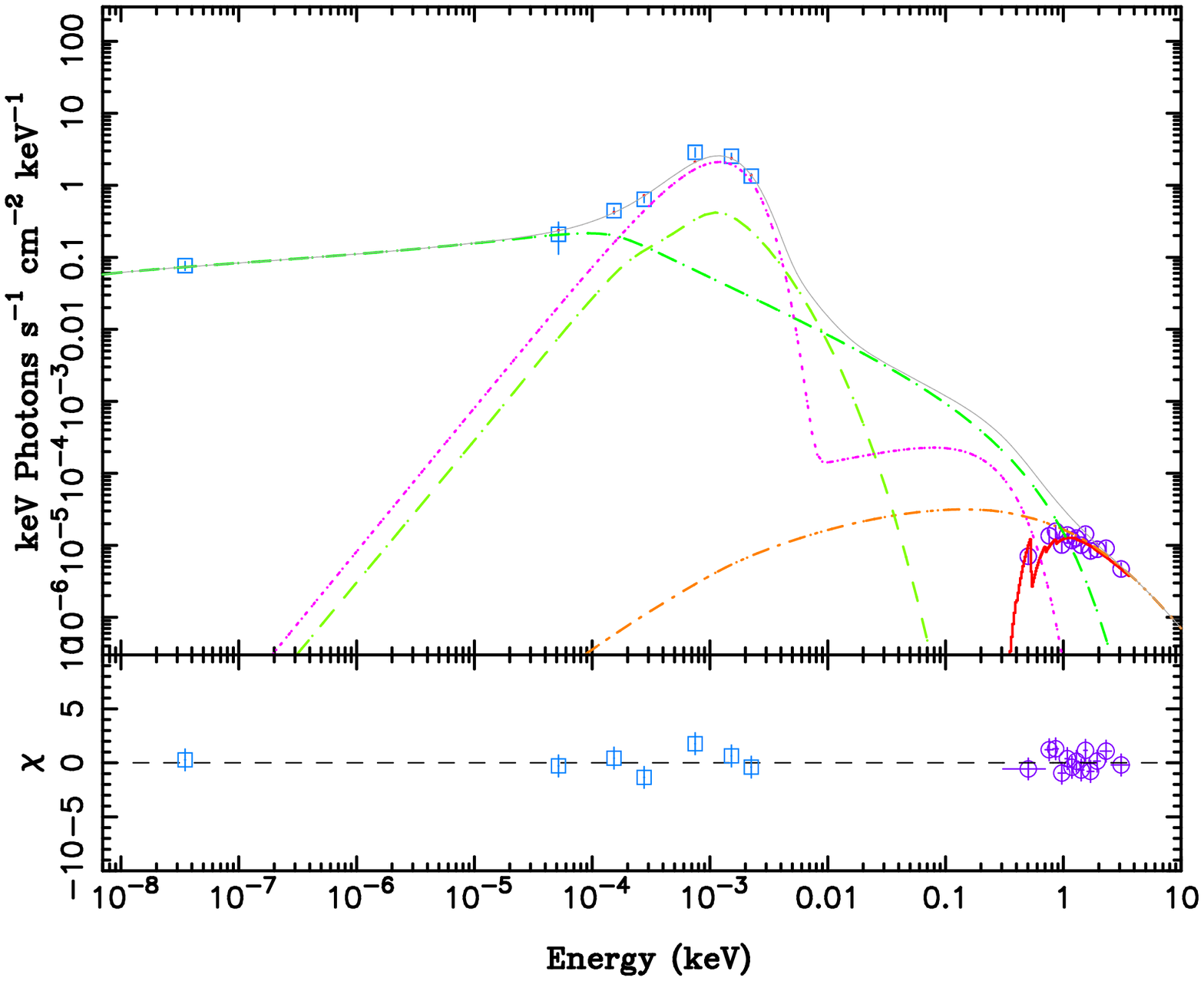}
\vspace{0.4cm}
\caption{Jet model fit to broadband A0620-00 data with
residuals.   The symbols represent the data,
while the solid red line is the model fit in
detector space.  Other indicated components are
not convolved with the detector matrices nor do
they include absorption, and serve only to
illustrate how the various emission mechanisms
and regions contribute to the continuum.   Solid
(gray): total spectrum; Dot-long-dashed (light
green): pre-acceleration inner jet synchrotron
emission; Dot-long-dashed  (darker green):
post-acceleration outer jet synchrotron;
Dot-dash-dash (orange): Compton emission from
the inner jet, including external disk photons
as well as synchrotron  self-Compton;
Dot-short-dash (magenta): thermal
multicolor-blackbody disk model plus single
blackbody representing the star. See Table~\ref{tab:sera} for the fitted
parameters. 
\label{fig:sera}}
\end{figure}

The model is most sensitive to the fitted parameter $N_{\rm j}$, which
acts as a normalization, though it is not strictly equivalent to the
total power in the jets (see discussion in
MNW05). It dictates the power initially divided
between the particles and magnetic field at the base of the jet, and
is expressed in terms of a fraction of 
$L_{\rm Edd}$.
Once $N_{\rm j}$ is specified and conservation is assumed, the macroscopic
physical parameters along the jet are determined assuming that the
jet power is roughly shared between the internal and external
pressures. 
The radiating particles enter the base of the jet where
the bulk velocities are lowest, with a quasi-thermal
distribution. Starting at location $z_{\rm acc}$ in the jets, a free
parameter, a fraction 85$\%$ of the particles are accelerated into a powerlaw
with index $p$, also a fitted parameter.  
The maximum energy of the accelerated leptons is calculated by setting
the acceleration rate to the local cooling rates from synchrotron and
inverse Compton radiation at $z_{\rm acc}$.  If the acceleration
process is diffusive Fermi acceleration, the acceleration rate depends
on the factor $f=\frac{(u_{\rm acc}/c)^2}{f_{sc}}$, where $u_{\rm acc}$ is
the shock speed relative to the bulk plasma flow, and $f_{\rm sc}$ is
the ratio of the scattering mean free path to the gyro-radius.  Because
neither plasma parameter is known, we fit for their combined
contribution via $f$, which thus reflects the {\it efficiency of
acceleration}. 
The particles in the jet radiatively cool via adiabatic expansion, the
synchrotron process, and inverse Compton up-scattering; however,
adiabatic expansion is assumed to dominate the observed effects of
cooling. A weak thermal accretion disk is assumed to be present, with an inner
disk temperature (somewhat arbitrarily) fixed at 
$T=10^6$ K, or $\sim90$ eV (inner disk temperatures between 50--200 keV are
typically obtained for higher Eddington ratio sources).  
This component is also included in the
Figure~\ref{fig:sera} and its photons are considered for local inverse Compton
up-scattering.  However they are negligible compared to the photons
produced by synchrotron radiation.  
The other main model parameters are the electron temperature $T_e$,
and the equipartition parameter between the magnetic field and the 
radiating (lepton) particle energy densities, $k$.  
A blackbody with temperature $4900$ K, consistent with the companion
star (Casares \etal 1993), is added to the model to account for the
optical emission.  An additional blackbody component has been also 
added to the fit, with normalization free to vary, in order to account
for possible contribution from the outer disk.  These photons are also
included in the Comptonization.  The ratio of the
`nozzle' (i.e. the pre-acceleration region) length to its radius has been
fixed to $1.5$, based on results in MNW05. The inclination angle
between the jet axis and line of sight $i$ has been fixed to
$43^\circ$, the mass fixed to 9.7 $M_\odot$ and the distance to 1.2
kpc, according to the recent results by Froning \etal (2007).  We wish
to stress that adopting the system parameters inferred by Gelino \etal
(2001) --i.e. 11 \msun for the BH mass and $i$=40.75$^{\circ}$-- 
does not result in a substantial change of the fitted parameters.
Starting with parameter values
similar to those found in other hard state BHBs, we have obtained a
reasonable fit to the data, with $\chi^2$/d.o.f.=14.3/11. The best fit model is 
shown in Figure~\ref{fig:sera}, with parameters and 90\%
confidence error bars given in Table~\ref{tab:sera}.

\subsection{Comparison to hard state sources}

Most of the free parameters have landed in ranges which we are starting to
recognize as `typical' based on higher luminosity sources such as Cyg X-1 and
GX~339-4 (MNW05), GRO~J1655-40 (Migliari et al., submitted to ApJ) and the low
luminosity AGN M81* (Markoff et al., in prep.).  Interestingly, the two main
differences appear to be related to the acceleration and equipartition.  In
higher luminosity sources we have found ratios of magnetic energy density to
the energy densities in radiating particles on the order of $\sim 1-5$, while
here our best fit value actually favors a slight domination of the particle
energy over the magnetic field ($0.1<k<0.2$).  The low error bar was limited
by the value 0.1, and thus does not represent a complete exploration of the
parameter space. Nevertheless, exact equipartition appears to be ruled out,
pointing towards a change in energy distribution.

What is quite different compared to higher luminosity sources,
however, is the required high-energy cutoff in the optically thin
synchrotron component, and thus in the accelerated electron
population.  This is determined by the acceleration parameter $f$
compared to the local cooling rates.  We find $f$ to be around two
orders of magnitude lower for A0620--00 than in higher luminosity
sources.  Interestingly, the only other black hole we can study
currently with similarly weak accretion is Sgr A*, the Galactic Center
super-massive BH.
In fact, the jet model was first developed in simplified form by
Falcke \& Markoff (2000), with the aim to determine whether the same
kind of model that could explain the inverted radio spectrum of Sgr A*
could also account for the newly discovered X-ray emission (Baganoff \etal 2000) . They concluded
that the SED of Sgr A* does not require a power law of optically thin
synchrotron emission after the break from its flat/inverted radio
spectrum. Therefore, if the radiating particles have a power-law
distribution, it must be so steep as to be indistinguishable from a
Maxwellian in the optically thin regime, i.e. they must be only weakly
accelerated\footnote{In this framework, radically different particle
  distributions, such as power laws and Maxwellians, may result in
  similar fits as long as the characteristic particle energy (minimum
  and peak energy, respectively for the power law and the Maxwellian)
  is similar. See MNW05, Appendix. }.
Here we have shown that something similar, albeit less extreme, is occurring
in the quiescent BHB A0620--00; either scenario implies that acceleration in
the jets is absent or very inefficient at $10^{-9}-10^{-8}L_{\rm Edd}$.

\section{Summary}
\label{sum}

We compile the radio/IR/optical spectra of three quiescent BHBs: V404 Cyg, XTE
J1118+480 and A0620--00, for which we also present new optical SMARTS
observations. Re-analysis of the archival \spi MIPS data for these systems
yields systematically higher values for the statistical uncertainties related
to sky subtraction with respect to the standard $\sim 10\%$ value that is
typically quoted for bright point-like sources.  While our revised values for
the 24 \mic fluxes are still consistent with those given by MM06 at the
3$\sigma$ level, they allow for a different interpretation of the measured
mid-IR excess with respect to the tail of the donor star thermal component.
We suggest that non-thermal emission from a jet could be responsible for a
significant fraction (or all) of the measured excess mid-IR emission. While
this possibly may not rule out the presence of circumbinary material, we argue
that the radio/IR/optical spectra of the three BHBs under consideration do not
require -- in a statistical sense -- the presence of an additional thermal
component. A variability study could definitively address
the question on the origin of the mid-IR excess, as, contrary to non-thermal
jet emission, circumbinary disk emission is expected to be steady. 

If non-thermal emission from a partially self-absorbed outflow is
indeed responsible for the measured mid-IR excess, then the
synchrotron luminosity of the jet, even excluding optically thin
radiation from the base, exceeds the measured 2-10 keV luminosity by a
factor of a few in all three systems. In turn, the jet mechanical
power in quiescence is greater than the bolometric (0.1-100 keV) X-ray luminosity
by several ($\simgt 4$) orders of magnitude.

We proceed by focusing on A0620--00, the lowest Eddington-ratio BHB
with a known radio counterpart, and construct its quiescent SED by
adding VLA, {\it Spitzer}, SMARTS and \cxo 
data.  In spite of the
non-simultaneity of the \spi observations with the radio/optical/X-ray
observations (which were taken over a two day period), we fit its
broadband SED of A0620-00 with a maximally jet-dominated model (MNW05).  
This is the first time that such a complex model is
applied in the context of quiescent BHBs, and with the strong
constraints on the jet break frequency cut-off provided by the \spi
data in the mid-IR regime.  In terms of best-fitting parameters, the major  
difference with respect to higher luminosity sources for which
this model has been successfully tested is in the value of the
acceleration parameter $f$ compared to the local cooling rates, which 
turns out to be two orders of magnitude lower for A0620--00.  This
weak acceleration scenario is reminiscent of the Galactic Center
super-massive BH Sgr A*. Within the jet model working hypothesis, 
both SEDs are in fact consistent with the hard X-ray emission
stemming primarily from inverse Compton processes in a corona/jet base
which is dominated by quasi-thermal particles.

\acknowledgments E.G. is funded by NASA through \cxo Postdoctoral
Fellowship grant number PF5-60037, issued by the \cxo X-Ray Center,
which is operated by the Smithsonian Astrophysical Observatory for
NASA under contract NAS8-03060. J.A.T. acknowledges partial support
from \spi contract number 1278068. C.D.B. is funded by NSF grant
AST-0407063. S.B. acknowledges support by the Ing.~Aldo Gini
Foundation. We are grateful to Mike Nowak for providing us with the
analysis scripts for ISIS.


\clearpage


\begin{center}
\newcommand\tabspace{\noalign{\vspace*{0.7mm}}}
\def\errtwo#1#2#3{$#1^{+#2}_{-#3}$}

\begin{deluxetable}{clrrr} 
\setlength{\tabcolsep}{0.15in} 
\tabletypesize{\scriptsize} 
\tablewidth{0pt} 
\tablecaption{Spitzer observations of quiescent black hole binaries\label{tab:spi}.}
\tablehead{ \colhead{Target}
           &  
           &         
           & \colhead{Flux ($\mu$Jy)}   
           & 
           \\ 
           & Method
           & $4.5 \mu$m
           & $8.0 \mu$m
	   & $24.0 \mu$m
	            } 
\startdata 
	    V404 Cyg
            & Ap. Photometry      
            & 3336
            & 1820
	    & 414$\pm$220
	\\
	    & PRF fitting
	    & 3220
	    & 1760
	    & 436$\pm$220
	\\      
 	    XTE J1118+480
            & Ap. Photometry     
            & 69
            & 59
	    & $<$50
	\\
	    & PRF fitting
	    & 69
	    & 58
	    & $<$50
            \\ 
	A0620--00
            & Ap. Photometry       
            & 412 
            & 288  
	    & 138$\pm$65 
	\\
	    & PRF fitting
	    & 380
	    & 305
	    &121$\pm$65 
            \\   
\tabspace 
\enddata 
 
\tablecomments{Un-dereddened values. Unless otherwise noted, flux
  errors are taken to be 10$\%$, due to calibration systematic
  errors. For the extinction corrections, we used the following
  values: V404 Cyg: $A_V$=2.8 (Shahbaz \etal 2003); XTE J1118+480: $A_V$=0.06
  (Gelino \etal 2006); A0620--00: $A_V$=1.2 (Wu \etal 1983). }
 
\end{deluxetable} 

\end{center}

\clearpage


\begin{center}
\newcommand\tabspace{\noalign{\vspace*{0.7mm}}}
\def\errtwo#1#2#3{$#1^{+#2}_{-#3}$}

\begin{deluxetable}{ccccc} 
\setlength{\tabcolsep}{0.07in} 
\tabletypesize{\scriptsize} 
\tablewidth{0pt} 
\tablecaption{Single blackbody fits to the IR-optical spectra\label{tab:bb}}
\tablehead{  \colhead{Target}
           & \colhead{$R/D$} 
           & \colhead{$T_{\rm fit}$}       
           & \colhead{$T_{\rm star}$}
	   & \colhead{$\chi^2/$d.o.f.} 
           \\ 
           \colhead{(1)}
           & \colhead{(2)} 
           & \colhead{(3)}       
           & \colhead{(4)}
	   & \colhead{(5)}
          } 
\startdata 
            V404 Cyg
	    & 5.4$\pm$0.2
            & 4533$\pm$80
            & 5500
	    &10.3/7
            \\ 
	    XTE J1118+480
	    & 0.7$\pm$0.1
	    & 3850$\pm$113
	    & 4250
	    & 69.6/5
		\\
       	  A0620--00
	    & 2.0$\pm$0.1
	    & 4468$\pm$104
	    & 4900
	    & 21.8/4
            \\
\tabspace 
\enddata \tablecomments{Columns are: (1) Source name; (2) Fitted star radius over distance,
  $R/D$, times $10^{-11}$; (3) Fitted star temperature, $T_{\rm fit}$, in $K$;
  (4) Star temperatures as found in the literature, $T_{\rm star}$, in $K$
  (references for the star temperature and system inclination and distance are
  the same as listed in the caption of Table~\ref{tab:spi} for the extinction
  values); (5) Fitted $\chi^2$ over degrees of freedom (d.o.f.).}
 
\end{deluxetable} 

\end{center}

\clearpage


\begin{center}
\newcommand\tabspace{\noalign{\vspace*{0.7mm}}}
\def\errtwo#1#2#3{$#1^{+#2}_{-#3}$}

\begin{deluxetable}{cccccc} 
\setlength{\tabcolsep}{0.07in} 
\tabletypesize{\scriptsize} 
\tablewidth{0pt} 
\tablecaption{Double blackbody fits to the IR-optical spectra\label{tab:bb-bb}}
\tablehead{  \colhead{Target}
           & \colhead{$(R/D)_1$} 
           & \colhead{$T_{\rm fit,1}$}
           & \colhead{$(R/D)_2$} 
	   & \colhead{$T_{\rm fit,2}$}          
	   & \colhead{$\chi^2/$d.o.f.} 
           \\ 
           \colhead{(1)}
           & \colhead{(2)} 
           & \colhead{(3)}
	   & \colhead{(4)} 
           & \colhead{(5)}
	   & \colhead{(6)}
          } 
\startdata 
            V404 Cyg
	    & 5.09$\pm$0.02
            & 4623$\pm$94
            & 30$\pm$18
            & 489$\pm$169
	    &1.4/5
            \\ 
	    XTE J1118+480
	    & 0.55$\pm$0.04   
	    & 4234$\pm$150
	    & 4$\pm$1
	    & 754$\pm$140
	    & 12.0/3
	\\
		    A0620--00
	    & 1.7$\pm$0.1
	    & 4691$\pm$149
	    & 23$\pm$10
	    & 393$\pm$83
	    & 2.0/2
            \\
\tabspace 
\enddata \tablecomments{Columns are: (1) Source name; (2)\&(4) Fitted blackbody radius over distance, times $10^{-11}$; (3)\&(5) Fitted blackbody temperature, in $K$; (6) Reduced $\chi^2$. Subscripts 1 and 2 indicate the first and
  second blackbody components.}
 
\end{deluxetable} 

\end{center}

\clearpage


\begin{center}
\newcommand\tabspace{\noalign{\vspace*{0.7mm}}}
\def\errtwo#1#2#3{$#1^{+#2}_{-#3}$}

\begin{deluxetable}{cccccccc}
\setlength{\tabcolsep}{0.07in} 
\tabletypesize{\scriptsize} 
\tablewidth{0pt} 
\tablecaption{Blackbody + broken power law fits to the
radio-IR-optical spectra. 
\label{tab:bb-pl}}
\tablehead{  \colhead{Target}
           & \colhead{$(R/D)$} 
           & \colhead{$T_{\rm fit}$}
           &  \colhead{$F_{\nu_0}$}
	   & \colhead{$\alpha_1$}
    	   & \colhead{$\chi^2$/d.o.f.}
           \\ 
	    \colhead{(1)}
           & \colhead{(2)} 
           & \colhead{(3)}
	   & \colhead{(4)} 
           & \colhead{(5)} 
	   &  \colhead{(6)} 
          } 
\startdata 
            V404 Cyg
	    & 5.0$\pm$0.2
	    & 4626$\pm$94
	    & 448$\pm$189
	    & 0.02$\pm$0.04
            & 4.0/9
   \\ 
	    XTE J1118+480
	    & 0.5$\pm$0.1
	    & 4302$\pm$211
	    & 62$\pm$23
	    & 0.27$\pm$0.39
            & 6.2/3
	\\
	  A0620--00
	    & 1.54$\pm$0.03
	    & 4897$\pm$6
	    & 148$\pm$1
	    & 0.113$\pm$0.001
            & 7.8/3
            \\
\tabspace 
\enddata 

\tablecomments{Columns are: (1), (2), (3): see Table 1; (4) Fitted power law
normalization at $\nu_0=10^{14}$ Hz, in $\mu$Jy (the broken power-law
expression is given in equation~\ref{eq:bp}; we fixed $\nu_b=10^{14}$ Hz and
$\alpha_2=-0.8$); (5) Fitted power law index below $\nu_b$; (6) Reduced
$\chi^2$.  }

\end{deluxetable} 

\end{center}

\clearpage


\begin{center}
\newcommand\tabspace{\noalign{\vspace*{0.7mm}}}
\def\errtwo#1#2#3{$#1^{+#2}_{-#3}$}

\begin{deluxetable}{cccc} 
\setlength{\tabcolsep}{0.07in} 
\tabletypesize{\scriptsize} 
\tablewidth{0pt} 
\tablecaption{A0620--00: SMARTS observations\label{tab:smarts}}
\tablehead{ \colhead{Band}
           & \colhead{UT start} 
           & \colhead{mag$^{a}$}       
           & \colhead{Flux$^{b}$ ($\mu$Jy)} 
           \\ 
          } 
\startdata 
            $V$
	    & 05Aug18-09:28:15 
            & 17.75$\pm$0.03
            & 884$\pm$68
            \\ 
	    $I$
	    & 05Aug18-09:19:31
	    & 16.04$\pm$0.05
	    & 1673$\pm$161
            \\
	    $H$
	    & 05Aug18-09:28:12
	    & 14.6$\pm$0.1
	    & 1910$\pm$283
	\\
\tabspace 
\enddata 
 
\tablecomments{$^{a}$Un-dereddened values.\\$^{b}$De-reddened values (adopting $A_V$=1.2),
allowing for an extra 0.05 mag uncertainty due to systematic calibration
errors.}  
 
\end{deluxetable} 

\end{center}

\clearpage


\begin{center}
\newcommand\tabspace{\noalign{\vspace*{0.7mm}}}
\def\errtwo#1#2#3{$#1^{+#2}_{-#3}$}

\begin{deluxetable}{ccccc} 
\setlength{\tabcolsep}{0.07in} 
\tabletypesize{\scriptsize} 
\tablewidth{0pt} 
\tablecaption{Jet power\label{tab:jets}}
\tablehead{ \colhead{Target}
           & \colhead{$\alpha_1$} 
           & \colhead{D}       
           & \colhead{$L_{\rm j,tot}$}
   	   & \colhead{$L_{\rm j,rad}/L_X$}
           \\ 
	   \colhead{(1)}     
           &   \colhead{(2)}     
           & \colhead{(3)}       
           & \colhead{(4)}
   	   &  \colhead{(5)}
           \\ 
          } 
\startdata 
            V404 Cyg
	    & 0.022 
            & 4 
            & $>$1.7$\times 10^{34}$
	    & 2.8
            \\ 
	    XTE J1118+480
	    & 0.270
            & 1.8
            & $>$3.7$\times 10^{32}$
	    & 5.4
	\\
	    A0620--00
	    & 0.113
            & 1.2
            & $>$4.5$\times 10^{32}$
	    & 3.8
	\\
\tabspace 
\enddata 
 
\tablecomments{Columns are: (1) Source name; (2) Fitted jet spectral
index below $\nu_b=10^{14}$ Hz; (3) Distance, in kpc; (4) {\it Total}
(kinetic + radiative) jet power, in erg s$^{-1}$; (5) Ratio between
the {\it radiative} jet power, integrated up to $\nu_b$, and the
quiescent X-ray luminosity $L_{\rm X}$, between 2--10 keV. $L_{\rm
j,tot}$ is calculated assuming no Doppler boosting, and a
(conservative) 5$\%$ radiative efficiency; as such, it represents a
strict lower limit to the total jet power.  Accordingly, $L_{\rm
j,rad}=0.05\times L_{\rm j,tot}$ only accounts for the partially
self-absorbed synchrotron emission from the jet. Quiescent X-ray
luminosities are taken from: V404 Cyg: Garcia et al. (2001), Kong et
al. (2002), Hynes et al. (2004).  XTE J1118+480: McClintock \etal
(2004). A0620--00: Kong \etal (2002), Gallo \etal (2006).}

\end{deluxetable} 
\end{center}

\clearpage


\begin{center}
\newcommand\tabspace{\noalign{\vspace*{0.7mm}}}
\def\errtwo#1#2#3{$#1^{+#2}_{-#3}$}

\begin{deluxetable}{ccccccccc} 
\setlength{\tabcolsep}{0.09in} 
\tabletypesize{\scriptsize} 
\tablewidth{0pt} 
\tablecaption{Jet Model for A0620--00\label{tab:sera}} 
\tablehead{ \colhead{$N_{\rm H}$}
           & \colhead{$N_{\rm j}$} 
           & \colhead{$r_{0}$}       
           & \colhead{$z_{\rm acc}$} 
           & \colhead{$T_{\rm e}$}          
           & \colhead{$p$}   
           & \colhead{$f$}
           & \colhead{$k$}            
           & \colhead{$BB_{\rm norm}$}
           \\ 
            (1)
           & (2)  
           & (3)  
           & (4)  
           & (5)
           & (6)
           & (7)
	   & (8)
	   & (9) 
          } 
\startdata 
             \errtwo{3.6}{0.7}{1.1} 
             & \errtwo{14.6}{0.4}{7.3} 
             & \errtwo{3.9}{2.2}{0.1}  
             & \errtwo{25}{272}{2}      
             & \errtwo{2.57}{0.29}{0.01} 
             & \errtwo{2.5}{0.1}{0.3} 
             & \errtwo{5.1}{1.0}{1.5} 
             & \errtwo{0.1}{0.1}{0.0} 
 	     & \errtwo{0.5}{0.2}{0.1} 
            \\ 
\tabspace 
\enddata 
 
\tablecomments{Columns are: fitted (1) Hydrogen equivalent column density, in
  $10^{22}$ cm$^{-2}$; (2) Model internal normalization, expressed in units of
  $10^{-4}L_{\rm Edd}$: it dictates the power dived by the particles and the
  magnetic field at the base; (3) Jet base (or `nozzle') radius, in
  units of gravitational radii $r_g=GM_{\rm BH}/c^2$; (4) Acceleration
  region, $z_{\rm acc}$, in $r_g$; it sets the location along the jet at
  which (a fraction of) the particles start being accelerated; (5) Temperature
  of the relativistic quasi-Maxwellian distribution with which the leptons
  enter the jet, in $10^{10}$ K; (6) Power law index of the accelerated
  electron distribution, $p$, where $N(E)\propto E^{-p}$; (7) Acceleration
  parameter, $f$, in units of $10^{-6}$: sets the balance between particle
  acceleration and radiative plus adiabatic cooling, such that the
  quasi-thermal particles be energized into a power-law tail; (8)
  Equipartition parameter, $k=(u_{\rm B }/u_{\rm rad}$): the ratio between the
  energy density in radiating leptons and the magnetic field energy density;
  (9) Internal disk blackbody normalization, in $10^{30}$ erg s$^{-1}$. We
  fixed the BH mass, distance and inclination of A0620--00 to: 9.7
  $M_{\odot}$, 1.2 kpc and 43$^{\circ}$ (Froning \etal 2007), yielding
  $\chi^2_{\rm red}=1.3$. Similar parameters, within the errors, are obtained
  adopting a mass of 11 $M_{\odot}$ and inclination of 40.75$^{\circ}$ (Gelino
  \etal 2001). Error bars are given at the 90\% confidence level.}
 
\end{deluxetable} 

\end{center}


\clearpage


\clearpage


\clearpage


\clearpage


\begin{thebibliography}{}




\bibitem[]{}
Baganoff F. \etal 2000, Bulletin of the American Astronomical Society, 32, 1184 

\bibitem{bb99}
Blandford R. \& Begelman C., 1999, MNRAS, 303, L1


\bibitem{bradley} 
Bradley C., Hynes R., Kong A., Haswell C., Casares J., Gallo E., 2007, ApJ, in
press (arXiv:astro-ph/07062652v1)






\bibitem[]{}
Cardelli J., Clayton, G., Mathis, J., 1989, ApJ, 345, 245

\bibitem[]{}
Casares J., Charles P. A., Naylor T., Pavlekno E. P., 1993, MNRAS, 265, 834



\bibitem{corbel06}
Corbel S., Tomsick J., Kaaret P., 2006, ApJ, 636,


\bibitem{corbel03}
Corbel S., Nowak M., Fender R. P., Tzioumis A. K., Markoff S., 2003, A\&A,
400, 1007

\bibitem[]{}
Corbel S. \& Fender R., 2002, ApJ, 573, L35






\bibitem[]{}
Dale D. \etal 2005,  ApJ, 633, 857






\bibitem[]{}
Falcke H. \& Markoff S., 2000, A\&A, 362, 113 
 	
\bibitem{fb95}
Falcke H. \& Biermann P. L., 1996, A\&A, 308, 321

\bibitem[]{}
Fazio G. \etal 2004, ApJS, 154, 10

\bibitem{fender06}
Fender R. P., 2006, in Lewin W. H. G., van der Klis M., eds, Compact
Stellar X-Ray Sources. Cambridge Univ. Press, Cambridge 

\bibitem{fbg}
Fender R. P., Belloni T., Gallo E., 2004, MNRAS, 355, 1105



\bibitem{fgj}
Fender R. P., Gallo E., Jonker P. G., 2003, MNRAS, 343, L99
	

\bibitem{fender01}
Fender R. P., 2001, MNRAS, 322, 31




\bibitem[]{}
Froning C., Robinson E., Bitner M., 2007, ApJ, in press (arXiv:astro-ph/07040267v1)



\bibitem{g07}
Gallo E., 2007, in Proc. of `The Multicolored Landscape of Compact Objects and
their Explosive Origins', AIP Conf. Proc. (arXiv:astro-ph/0702126v1)

\bibitem{gallo06}
Gallo E. \etal 2006, MNRAS, 370, 1351

\bibitem{g05a}
Gallo E., Fender R. P., Hynes R. I., 2005b, MNRAS, 356, 1017

\bibitem{nat}
Gallo E., Fender R., Kaiser, C., Russell, D., Morganti, R., Oosterloo, T., Heinz, S., 2005a, Nature, 436, 819



\bibitem{gfp}
Gallo E., Fender R. P., Pooley G. G., 2003, MNRAS, 344, 60

\bibitem[]{}
Gelino D., Balman S., Kiziloglu \"U., Yilmaz A., Kalemci E., Tomsick J., 2006, ApJ, 642, 438

\bibitem{gelino01}
Gelino D. M., Harrison T. E., Orosz J. E. 2001, AJ, 122, 2668






\bibitem[]{}
Heinz S. \& Sunyaev R., 2003, MNRAS, 343, L59

\bibitem{heinzgrimm}
Heinz S. \& Grimm H.-J. 2005, ApJ, 633, 384


\bibitem{hj00}
Hjellming P. M., Rupen M. P., Mioduszewski A. J., Narayan R., 2000, ATel 54


\bibitem{ho2}
Homan J., Buxton, M., Markoff, S., Bailyn, C., Nespoli, E., Belloni, T., 2005, ApJ, 624, 295

\bibitem{ho1}
Homan J. \& Belloni T., 2005, A\&SS, 300, 107

\bibitem[]{}
Houck J. \& De Nicola L., 2000, in ASP Conf. Ser., Vol. 216, Astronomical Data Analysis Software and Systems IX, eds. N. Manset, C. Veillet, D. Crabtree (San Francisco: ASP), 591 

\bibitem{hynes06}
Hynes R. \etal 2006, ApJ, 651, 401

\bibitem{hynes04}
Hynes R. \etal 2004, ApJ, 611, L125

\bibitem[]{}
Hynes R. I. \etal 2003, MNRAS, 345, 292







\bibitem[]{}
K\"ording E., Fender R., Migliari S., 2006, MNRAS, 369, 1451

	
\bibitem[]{}
Makovoz D. \& Marleau F., 2005, PASP, 117, 1113

\bibitem{mnw05}
Markoff S., Nowak M. A., Wilms J., 2005, ApJ, 635, 1203 (MNW05)

\bibitem{mn04}
Markoff S. \& Nowak M. A., 2004, ApJ, 609, 972

\bibitem{markoff03}
Markoff S., Nowak, M., Corbel, S., Fender, R., Falcke, H., 2003, A\&A, 397, 645

\bibitem{mff}
Markoff S., Falcke H., Fender R., 2001, A\&A, 372, L25







\bibitem[]{}
Miller J., Homan, J., Steeghs, D., Rupen, M., Hunstead, R. W., Wijnands, R., Charles, P. A., Fabian, A. C., 2006a, ApJ, 653, 525

\bibitem[]{}
Miller J., Homan, J., Miniutti G., 2006b, ApJ, 652, L113

\bibitem{mccr06}
McClintock J. E., Remillard R. A., 2006, in Lewin W. H. G., van der Klis M.,
eds, Compact Stellar X-Ray Sources. Cambridge Univ. Press, Cambridge

\bibitem{mcc03}
McClintock J. E. \etal 2003, ApJ, 593, 435


\bibitem[]{}
Migliari S., Tomsick, J., Maccarone, T., Gallo, E., Fender, R., Nelemans, G., Russell, D., 2006, ApJ, 643, L41


	




\bibitem[]{}
Mirabel I. F., Dhawan V., Mignani R., Rodrigues I., Giglielmetti F., 2001, Nature, 413, 139


	

\bibitem{mm06}
Muno M. \& Mauerhan J., 2006, ApJ, 648, L135 (MM06)

\bibitem[]{}
Nowak, M.; Wilms, J., Heinz, S., Pooley, G., Pottschmidt, K., Corbel, S., 2005, ApJ, 629, 1006




\bibitem{ny94}
Narayan R., Yi I. 1994, ApJ, 428, L13









\bibitem[]{}
Rieke G. \etal 2004, ApJS, 154, 25

\bibitem{r07}
Russell D., Fender R., Gallo E., Kaiser C., 2007, MNRAS, 376, 1341

\bibitem{r06} 
Russell, D., Fender, R., Hynes, R., Brocksopp, C., Homan, J., Jonker, P., Buxton, M., 2006, MNRAS, 371, 1334 (R06)

\bibitem[]{}
Rykoff E., Miller J., Steeghs D., Torres M., 2007, submitted to ApJ (arXiv:astro-ph/0703497v1)

\bibitem{shahbaz03}
Shahbaz T., Dhillon V. S., Marsh T. R., Zurita C., Haswell C. A., Charles
P. A., Hynes R. I., Casares J., 2003, MNRAS, 346, 1116










\bibitem[]{}
Taam R. \& Spruit H., 2001, ApJ, 561, 329

\bibitem[]{}
Wang Z., Chakrabarty D., Kaplan D., 2006, Nature, 440, 772

\bibitem[]{}
Wu C.-C., Aalders J., van Duinen R., Kester D., Wesselius P., 1976, A\&A, 50, 445

\bibitem{xcui}
Xue Y. \& Cui W., 2007, A\&A, 466, 1053

\bibitem{yuan}
Yuan F., Cui W., Narayan R., 2005, ApJ, 620, 905

\bibitem[]{}
Zurita C., Casares J., Hynes R., Shahbaz T., Charles P., Pavlenko E., 2004, MNRAS, 352, 877

\bibitem[]{}
Zurita C., Casares J., Shahbaz T., 2003, ApJ, 582, 369

\end{thebibliography}
\end{document}